\documentclass[useAMS,usenatbib]{mn2e}

\usepackage{lscape}
\usepackage{epsfig}
\usepackage{times}

\def\gtsim{\mathrel{\spose{\lower.5ex \hbox{$\mathchar"218$}}
     \raise.4ex\hbox{$\mathchar"13E$}}}
\def\ltsim{\mathrel{\spose{\lower.5ex\hbox{$\mathchar"218$}}
     \raise.4ex\hbox{$\mathchar"13C$}}}
\def\aFe{[$\alpha/{\rm Fe}$]}

\def\Hb{${\rm H}{\beta}$}

\def\Mgb{{\rm Mg}\,$_b$}
\def\Fe{$\langle {\rm Fe}\rangle$}

\def\MH{[$Z/{\rm H}$]}
\def\ZH{[$Z/{\rm H}$]}
\def\MgFe{[${\rm MgFe}$]$'$}

\def\Mgd{{\rm Mg}\,$_2$}

\def\Rnc{$r_{\rm{ d95}}$} 
\def\Rlast{$r_{\rm{ last}}$}

\def\agelum{$\langle\,{\rm Age}\,\rangle_L$} 
\def\metlum{$\langle\,[Z/{\rm H}]\,\rangle_L$} 
\def\agelumo{$\langle\,{\rm Age}\,\rangle_L^{\rm old}$} 
\def\metlumo{$\langle\,[Z/{\rm H}]\,\rangle_L^{\rm old}$} 
\def\agelumy{$\langle\,{\rm Age}\,\rangle_L^{\rm young}$} 
\def\metlumy{$\langle\,[Z/{\rm H}]\,\rangle_L^{\rm young}$} 

\def\kms{$\rm km\;s^{-1}$}

\def\apj{ApJ}
\def\aj{AJ}
\def\apjl{ApJL}
\def\apjs{ApJS}
\def\mnras{MNRAS}

\def\aap{A\&A}

\def\araa{ARA\&A}
\def\pasp{PASP}

\def\spose#1{\hbox to 0pt{#1\hss}}

\begin{document}

\title[Stellar populations in the discs of ten spiral galaxies]{Study of
  the stellar population properties in the discs of ten spiral galaxies}
\author[L. Morelli et al.]{L.~Morelli$^{1,2}$\thanks{E-mail: 
  lorenzo.morelli@unipd.it}, E.~M.~Corsini$^{1,2}$, A.~Pizzella$^{1,2}$, 
  E. Dalla Bont\`a$^{1,2}$, L.~Coccato$^{3}$ and
  \and J.~M\'endez-Abreu$^{4}$\\ 
$^1$ Dipartimento di Fisica e Astronomia ``G. Galilei'', Universit\`a
     di Padova, vicolo dell'Osservatorio 3, I-35122 Padova, Italy\\ 
$^2$ INAF--Osservatorio Astronomico di Padova,
     vicolo dell'Osservatorio 5, I-35122 Padova, Italy\\
$^3$ European Southern Observatory, Karl-Schwarzschild-Stra$\beta$e 
    2, D-85748 Garching bei M\"unchen, Germany\\
$^4$ School of Physics and Astronomy, University of St. Andrews, 
    SUPA, North Haugh, KY16 9SS St. Andrews, UK}
\date{{\it Draft version on \today}}

\maketitle


\begin{abstract}

We investigated the properties of the stellar populations in the discs
of a sample of ten spiral galaxies. Our analysis focused on the
galaxy region where the disc contributes more than 95 per cent of
total surface brightness in order to minimise the contamination of the
bulge and bar.

The luminosity-weighted age and metallicity were obtained by fitting
the galaxy spectra with a linear combination of stellar population
synthesis models, while the total overabundance of $\alpha$-elements
over iron was derived by measuring the line-strength indices.

Most of the sample discs display a bimodal age distribution and they
are characterised by a total \aFe\ enhancement ranging from solar and
supersolar.
We interpreted the age bimodality as due to the simultaneous presence
of both a young (Age$\,\leq\,4$ Gyr) and an old (Age$\,>\,$4 Gyr)
stellar population.  The old stellar component usually dominates the
disc surface brightness and its light contribution is almost constant
within the observed radial range. For this reason, no age gradient is
observed in half of the sample galaxies. The old component is slightly
more metal poor than the young one. The metallicity gradient is
negative and slightly positive in the old and young components,
respectively.

These results are in agreement with an inside-out scenario of disc
formation and suggest a reduced impact of the radial migration on the
stellar populations of the disc.  The young component could be the
result of a second burst of star formation in gas captured from the
environment.

\end{abstract}


\section{Introduction}
\label{sec:introduction}

Studying the properties of the stellar populations of galaxies is 
essential to give a comprehensive picture of their formation and
evolution. Indeed, all the processes driving the assembly history of
galaxies leave a fingerprint in the radial profiles of the luminosity weighted age,
metallicity, and \aFe\ enhancement of the stellar component. Thus, the
observed properties of the stellar populations can be used against the
predictions of theoretical models and numerical simulations.

Many studies focus on the colours of galaxies and their structural
components. The mean colours of the stellar populations change
smoothly from red to blue along the Hubble sequence. Moreover, the
galactic discs tend to be bluer than spheroids, with them being either elliptical
galaxies or bulges of lenticular and spiral galaxies
\citep{dejong96,taylor05,Driver2006}. However, photometric data alone
is not enough to distinguish whether these systematic differences are
an effect of age or metallicity then making impossible to drive conclusive
results about the stellar populations in galaxies beyond the Local
Group \citep{worthey94}.

\begin{table*}
\caption{Properties of the sample galaxies. The columns show the
  following: 1, galaxy name; 2, morphological classification from Lyon
  Extragalactic Database (LEDA); 3, Hubble type (LEDA); 4) apparent
  isophotal diameters measured at a surface-brightness level of $\mu_B
  = 25$ mag arcsec$^{-2}$ (LEDA); 5, total observed blue magnitude
  from LEDA; 6, radial velocity with respect to the cosmic microwave
  background reference frame (LEDA); 7, distance obtained as in
  \citet{moreetal08, pizzetal08, morelli2012} adopting $H_0 = 75$ km
  s$^{-1}$ Mpc$^{-1}$; 8, absolute total blue magnitude from $B_T$
  corrected for extinction as done in LEDA and adopting the distance
  $D$; 9, disc scalelength; 10, central values of the velocity
  dispersion; 11, Source of the photometric and spectroscopic data: 
  ($1=$ \citet{moreetal08}, $2=$ \citet{pizzetal08}, $3=$
  \citet{morelli2012}, $4=$ \citet{morelli15}). }
\begin{center}
\begin{small}
\begin{tabular}{lcr cr cccccc}
\hline
\noalign{\smallskip}
\multicolumn{1}{c}{Galaxy} &
\multicolumn{1}{c}{Type} &
\multicolumn{1}{c}{$T$} &
\multicolumn{1}{c}{$D_{25}\,\times\,d_{25}$} &
\multicolumn{1}{c}{$B_T$} &
\multicolumn{1}{c}{$V_{\rm CMB}$} &
\multicolumn{1}{c}{$D$} &
\multicolumn{1}{c}{$M_{B_T}$}&
\multicolumn{1}{c}{$h$} &
\multicolumn{1}{c}{$\sigma$}&
\multicolumn{1}{c}{Source} \\ 
\multicolumn{1}{c}{} &
\multicolumn{1}{c}{} &
\multicolumn{1}{c}{} &
\multicolumn{1}{c}{(arcmin)} &
\multicolumn{1}{c}{(mag)} &
\multicolumn{1}{c}{(\kms)} &
\multicolumn{1}{c}{(Mpc)} &
\multicolumn{1}{c}{(mag)}&
\multicolumn{1}{c}{(arcsec)}&
\multicolumn{1}{c}{(\kms)}&
\multicolumn{1}{c}{} \\
\multicolumn{1}{c}{(1)} &
\multicolumn{1}{c}{(2)} &
\multicolumn{1}{c}{(3)} &
\multicolumn{1}{c}{(4)} &
\multicolumn{1}{c}{(5)} &
\multicolumn{1}{c}{(6)} &
\multicolumn{1}{c}{(7)} &
\multicolumn{1}{c}{(8)} &
\multicolumn{1}{c}{(9)} &
\multicolumn{1}{c}{(10)} &
\multicolumn{1}{c}{(11)}\\
\noalign{\smallskip}
\hline
\noalign{\smallskip}  
ESO-LV~1890070 & SABb &$3.8$  & $3.0\times2.0$ &12.31 & 2981  & 37.5  & $-20.56$& 31.1 & 91     &  2,4\\
ESO-LV~2060140 & SABc &$5.0$  & $1.6\times0.8$ &14.89 & 4672  & 60.5  & $-19.01$& 17.8 & 54.3   &  3\\
ESO-LV~4000370 & SBc  &$5.9$  & $1.8\times0.9$ &14.47 & 2876  & 37.5  & $-18.40$& 22.4 & 42.0   &  3\\
ESO-LV~4500200 & SBbc & $4.1$ & $1.9\times1.6$ &13.03 & 2118  & 31.6  & $-19.47$& 17.3 & 112    &  2,4\\
ESO-LV~5140100 & SABc &$5.1$  & $2.5\times1.9$ &12.88 & 2888  & 40.4  & $-20.13$& 27.1 & 60     &  2,4\\
ESO-LV~5480440 & S0/a &$-0.9$ & $1.2\times0.5$ &14.26 & 1696  & 22.6  & $-17.51$& 10.1 & 63.8   &  1\\
IC  1993       & SABb &$3.0$  & $1.4\times0.8$ &12.52 & 1065  &  17.0 & $-18.63$& 21.9 & 182.8  &  1\\
NGC 1366       & S0   &$-2.2$ & $2.1\times0.9$ &12.80 & 1308  &  17.0 & $-18.63$& 12.9 & 175.8  &  1\\
NGC 7643       & Sab  &$2.0$  & $1.4\times0.8$ &14.12 & 3837  &  50.2 & $-19.38$& 11.1 & 116.9  &  1\\
PGC~37759      & Sc   &$6.0$  & $0.6\times0.4$ &15.89 & 14495 & 193.2 & $-20.54$& 7.20 & 64.4   &  3\\
\noalign{\smallskip}
\hline
\noalign{\medskip}
\end{tabular}
\end{small}
\label{tab:sample}
\end{center}
\end{table*}

Spectroscopic data allow us to highly reduce the age-metallicity
degeneracy by studying the absorption features which are connected to
the properties of the stellar populations. Coupling the measurements of
both the line-strength indices of the Lick system \citep{faberetal85}
and other systems with higher spectral resolution \citep{johaetal10,
  vazdekisetal2010} with the predictions of single stellar population
(SSPs) models \citep{thmabe03, thomasetal2011, vazdekisetal2010} is
the most widely used method to derive the age, metallicity, and
\aFe\ enhancement of unresolved stellar populations in
galaxies. However, most of targets are elliptical galaxies
\citep{mehletal03, sancetal06p, annietal07} and bulges of disc
galaxies \citep{mooretal06, jabletal07, moreetal08, morelli2012, seidel15} for
which the assumption of SSP is a good approximation.
To date, only a few papers focused on the stellar populations of
galactic discs have been published because of their low surface
brightness and the presence of intense nebular emission lines which
make the spectroscopic analysis quite difficult. In addition, the
discs of galaxies are a reservoir of molecular gas \citep{davis12}
feeding more than one single episode of star formation. Therefore, the
SSP approach can not be used in discs and multiple stellar populations
are required to correctly recover the star formation history and the
stellar population properties in galaxy discs
\citep{morelli13,gonzalezdelgado2014,mcdermid15}.

\citet{yoacdalc08} and \citet{macaetal09} detected the presence of an
old stellar population (8-10 Gyrs) in the disc-dominated region of 9
edge-on and 8 low inclined spirals, respectively. This result was
confirmed by \citet{sancetal11} and \citet{sancetal14} who studied 62
nearly face-on spirals without finding any significant difference
between the discs of unbarred and barred galaxies. Discs are
characterised by a large spread in metallicity spanning values between
\ZH=0 and \ZH=-0.2 \citep{yoacdalc08}. The age and metallicity radial
profiles derived by \citet{macaetal09}, \citet{sancetal11}, and
\citet{sancetal14} display shallow or negative metallicity gradients. The
\aFe\ enhancement was only measured by \citet{yoacdalc08}. They found
that the stars of both the thick and thin disc have a solar abundance
ratio.

Tracing the radial profiles of the age, metallicity, and
\aFe\ enhancement in the discs in several nearby galaxies will allow
to extract statistically significant conclusions and perform a
comparison with the Milky Way \citep{freeman02, yong06}, and the other
disc galaxies of the Local Group, \citep{worthey05, davidge07,
  cioni09, gogarten10}, in order to understand the mechanisms driving
the disc formation.

Galactic discs are believed to form immediately after a major merging
of gas dominated systems \citep{robertson2006}, through an inside-out
process with the inner parts forming first due to their lower angular
momentum and the outer parts with higher angular momentum forming
later \citep{brook2004, munoetal07}. The inside-out scenario was
confirmed by \citet{trujillo06} studying the disc truncations at high
$z$. Within this framework, age and metallicity gradients are expected
to be measured across galactic discs with the older
\citep{matteucci89, boissier99} and more metal rich stars
\citep{munoetal07,roskar08, prochaska11} confined in the inner
regions.
Nevertheless, this prediction is correct under the assumption that stars remain
in the same region of the disc where they formed. Recently it has been
shown the possibility that the stellar orbits change in time and stars
might move inward or outward across the disc. This effect is referred as
stellar migration and it has been investigated both with a theoretical
approach \citep{jenkins90, sellwood02} and using numerical simulations
\citep{roskar08, dimatteo13, kubryk13, minchev14}. The process of
radial migration is expected to mix the stellar populations and
flatten the metallicity gradient with time.

The paucity of data to address such a relevant topic makes worthwhile
any effort to measure the properties of the stellar populations in
galactic discs. Therefore, we decided to study the stellar populations
in the discs of a sample of 10 spiral galaxies for which the
structural and kinematical properties of the bulge and disc were
already known \citep{moreetal08, pizzetal08, morelli2012,
  morelli15}. The paper is organized as follows. We present the
  sample of galaxies in \S \ref{sec:sample} and investigate the
  properties of the stellar populations of their discs in \S
  \ref{sec:populations}.  Finally, we discuss our conclusions in \S
  \ref{sec:conclusions}.

\section{Sample selection}
\label{sec:sample}

The sample comprises 10 disc galaxies with morphological types ranging
from S0 to Sc whose properties, including size, magnitude, and
distance, are listed in Table~\ref{tab:sample}. The galaxies belong to
the sample of nearby lenticulars and spirals studied by
\citet{pizzetal08} and \citet{moreetal08, morelli2012, morelli15} who
measured the surface-brightness distribution from broad-band
photometry and the stellar and ionized-gas kinematics from long-slit
spectra. We refer to these aforementioned papers for the details about
the selection criteria, photometric decomposition, and measurements of
the stellar kinematics and Lick indices of the sample galaxies.

The properties of the stellar populations of the bulges of the sample
galaxies were already investigated \citep{moreetal08, morelli2012,
  morelli15}, and here we extend the analysis deriving the stellar
populations in the discs.

\begin{figure*}
 \includegraphics[angle=0.0,width=0.991\textwidth]{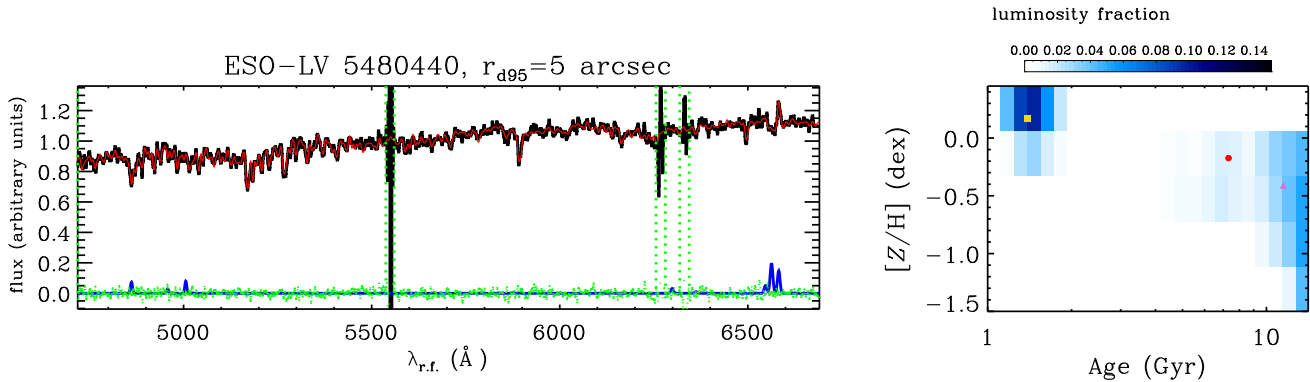}
 \caption{Left-hand panel: The rest-frame observed spectrum of
   ESO-LV~5480440 at \Rnc\ (black line) is plotted with the
   best-fitting model (red line) obtained as the linear combination of
   synthetic population models, ionized-gas emission lines (blue
   lines), and a low order multiplicative Legendre polynomial. The
   residuals (green dots) are calculated by subtracting the model from
   the observed spectrum. The vertical dashed lines correspond to
   spectral regions masked out in the fitting process. Right-hand panel:
   Age and metallicity distribution obtained from the best-fitting
   synthetic population models. The col or scale gives the luminosity
   fraction in each bin of age and metallicity. The red circle marks
   the luminosity-weighted averages of age (\agelum) and metallicity
   (\metlum). The yellow square and purple triangle refer to the
   luminosity-weighted averages of age and metallicity for the old
   ($\rm Age> 4$ Gyr) and young ($\rm Age\leq 4$ Gyr) components of
   the stellar population, respectively. }
 \label{fig:eso548}
\end{figure*}

\section{Stellar populations of the discs}
\label{sec:populations}

Measuring the stellar populations of the different galaxy components
from integrated spectra suffers from contamination. The properties
inferred for one component are indeed affected by those of the others
depending on their relative contribution to the galaxy surface
brightness. This issue is critical for bulges and bars, which are
always embedded in discs and overlap one to each other at small radii,
nonetheless it can be easily overcome for discs which dominate the
surface brightness distribution at large radii.

Since we were interested in investigating the stellar populations of
the disc component, we focused our analysis on the disc-dominated
region between \Rnc, which is the radius where the disc contributes
more than $95\%$ of the galaxy surface brightness, and \Rlast, which
is the farthest radius where the signal-to-noise ratio is sufficient
to measure the properties of the stellar populations in the available
spectra ($S/N>20$ per resolution element). 

For each galaxy, we adopted the photometric decomposition performed by
\citet{pizzetal08} and \citet{moreetal08, morelli2012, morelli15} to
define \Rnc. The structural parameters of the galaxies were derived
with Galaxy Surface Photometry Two-Dimensional Decomposition
\citep[{\sc GASP2D},][]{mendetal08} by assuming the observed
surface-brightness distribution to be the sum of a S\'ersic bulge, an
exponential disc, and, if necessary, a Ferrers bar
\citep{mendetal14}. The values of \Rnc\ and \Rlast\ for all the sample
galaxies are listed in Table~\ref{tab:results_1burst}.

\subsection{Age and metallicity}
\label{sec:age_metallicity}

The age and metallicity of the galactic components are crucial in
reconstructing the assembly history of galaxies, since the past events
of merging and star formation are imprinted in the stellar
populations.

\subsubsection{Fitting synthetic population models}
\label{sec:msp}

We analysed the galaxy spectra of \citet{pizzetal08} and
\citet{moreetal08, morelli2012, morelli15} to derive at \Rnc\ and
\Rlast\ the relative contribution of the stellar populations with
different age and metallicity to the observed surface-brightness
distribution.  As done by \citet{onodera12} and \citet{morelli13}, we
applied the Penalized Pixel Fitting code \citep[{\sc
    pPXF},][]{capems04}, including the Gas and Absorption Line Fitting
algorithm \citep[{\sc GANDALF},][]{sarzetal06} and a linear
regularization of the weights \citep{press92}, which were adjusted for
the sample spectra to deal with emission lines and derive both the
distribution of the luminosity fraction in different age and
metallicity bins, respectively.

We adopted 115 synthetic population models with Salpeter initial mass
function \citep{salp55}, age from 1 to 15 Gyr, and metallicity \MH\/
from $-1.5$ to $0.22$ dex. They were built from the stellar spectra
available in the Medium Resolution Isaac Netwon Telescope Library of
Empirical Spectra \citep[MILES,][]{sancetal06lib} with a spectral
resolution of ${\rm FWHM} = 2.54$ \AA\ \citep{beifiori11}. The
spectral resolution of the galaxy spectra was degraded to match that
of the synthetic population models. Then they were convolved with the
line-of-sight velocity distribution (LOSVD) obtained from the
available stellar kinematics and fitted to the galaxy spectrum using a
$\chi^2$ minimisation in pixel space. We simultaneously fitted the
galaxy spectra using emission lines in addition to the synthetic
population models. Only emission lines detected with a $S/N > 3$ were
taken into account. To make the fit result more sensitive to the
absorption lines of the galaxy spectrum than to its continuum shape,
we adopted a low order multiplicative Legendre polynomial to account
for both reddening and possible artifacts due to the flat fielding or
or flux calibration. Finally, we derived the stellar light fraction
within each bin of age and metallicity from the best-fitting synthetic
population models.
 
An example of the fitting procedure is shown in Fig.~\ref{fig:eso548}
for the spectrum at \Rnc\ of ESO-LV~5480440.  The age, metallicity, and
light fraction of the stellar populations at \Rnc\ and \Rlast\ of all
the sample galaxies are plotted in Fig.~\ref{fig:age_metallicity}.

\begin{figure*}
 \centering{ESO-LV~1890070\\}
 \includegraphics[angle=0.0,width=0.431\textwidth,height=0.24\textheight]{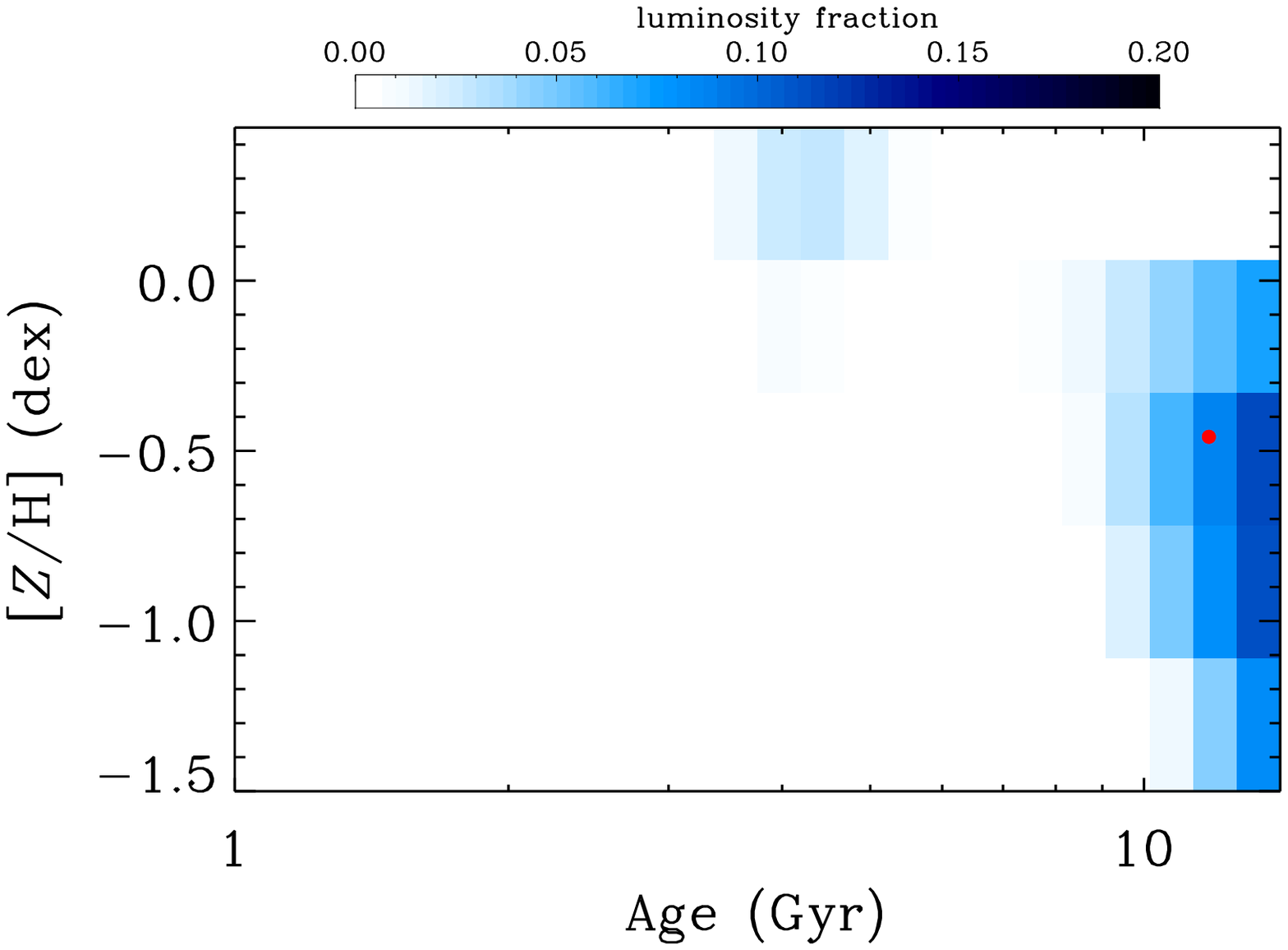}
 \includegraphics[angle=0.0,width=0.431\textwidth,height=0.24\textheight]{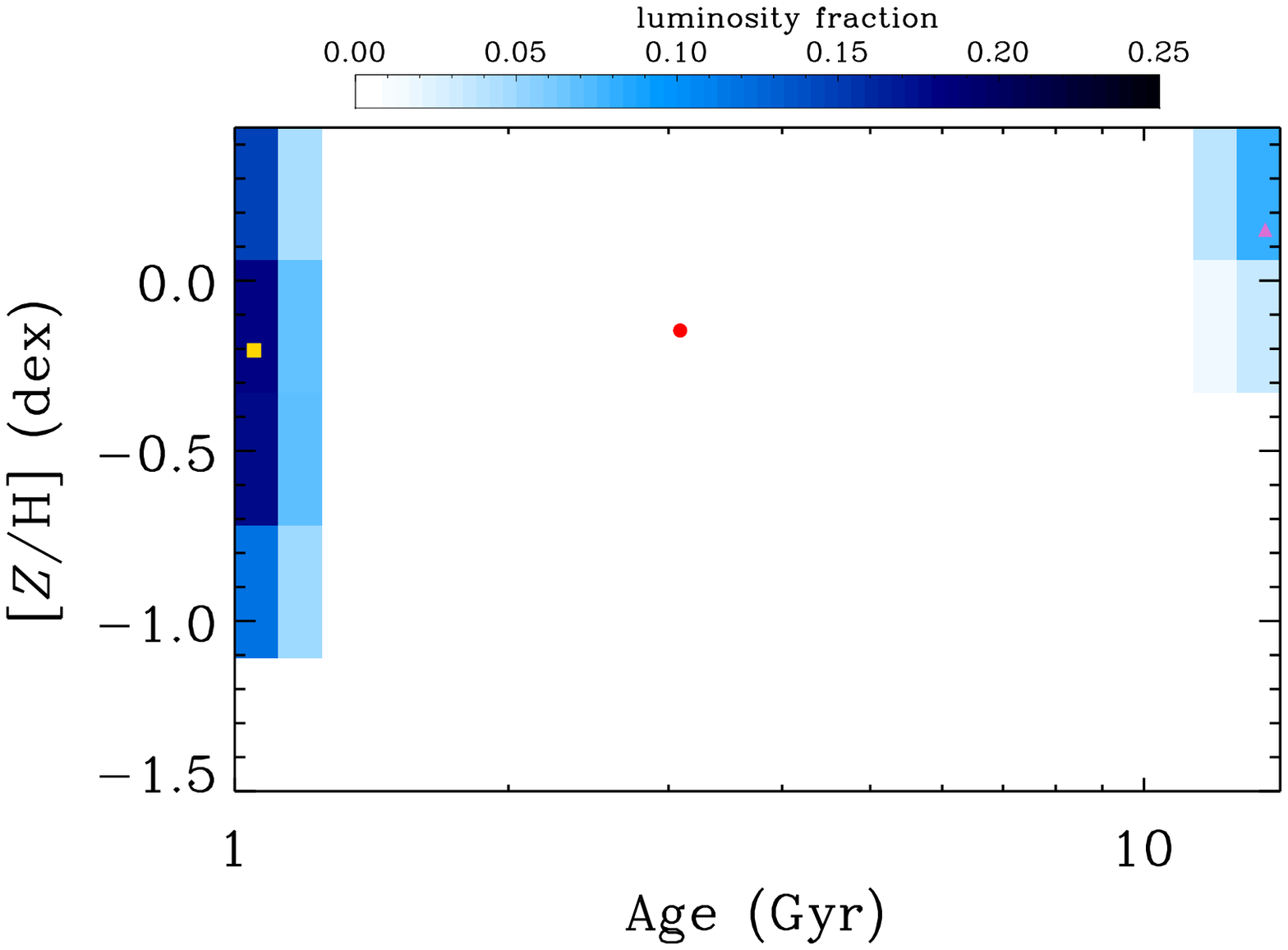}\\
 \centering{ESO-LV~2060140\\}
 \includegraphics[angle=0.0,width=0.431\textwidth,height=0.24\textheight]{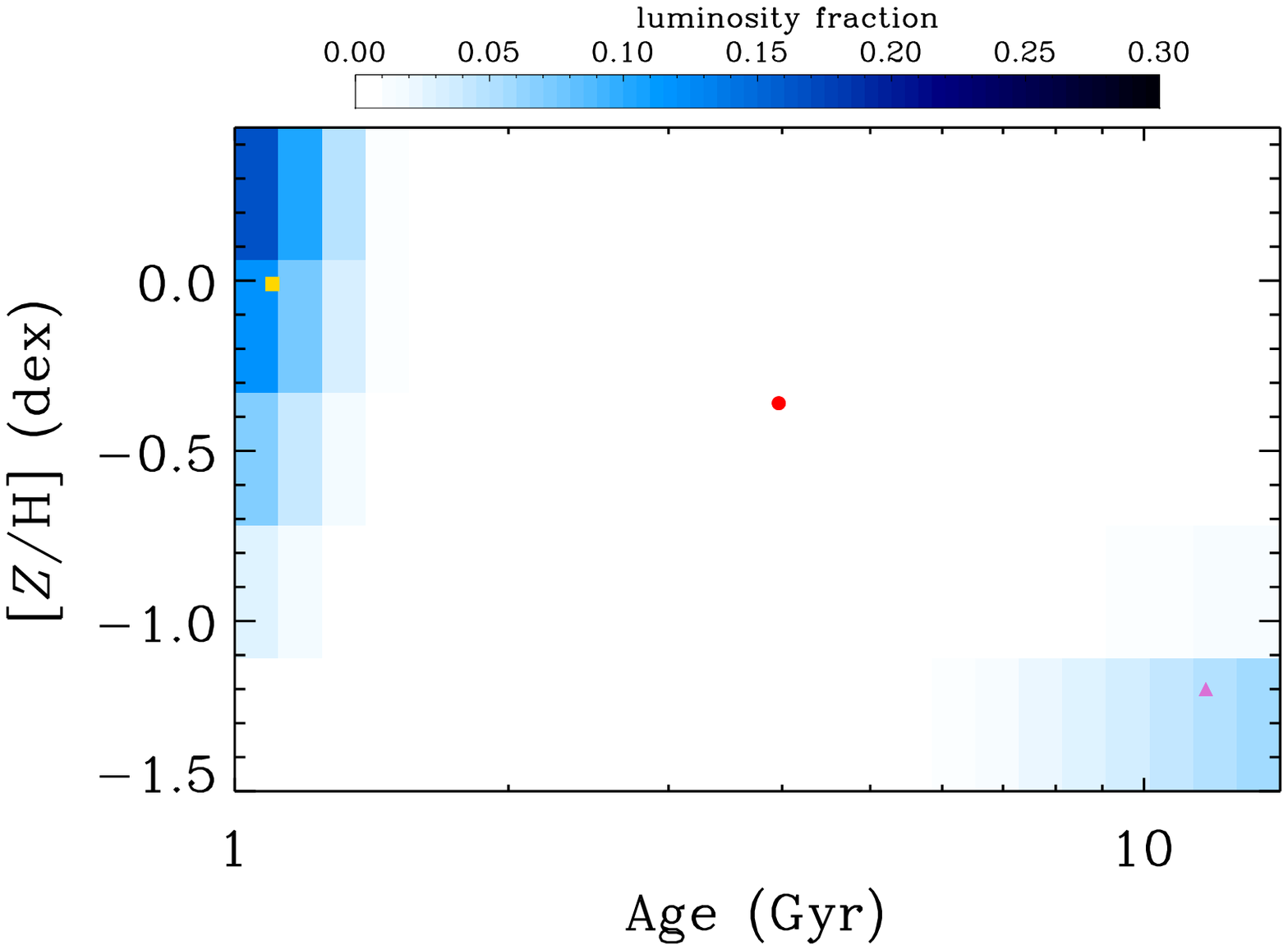}
 \includegraphics[angle=0.0,width=0.431\textwidth,height=0.24\textheight]{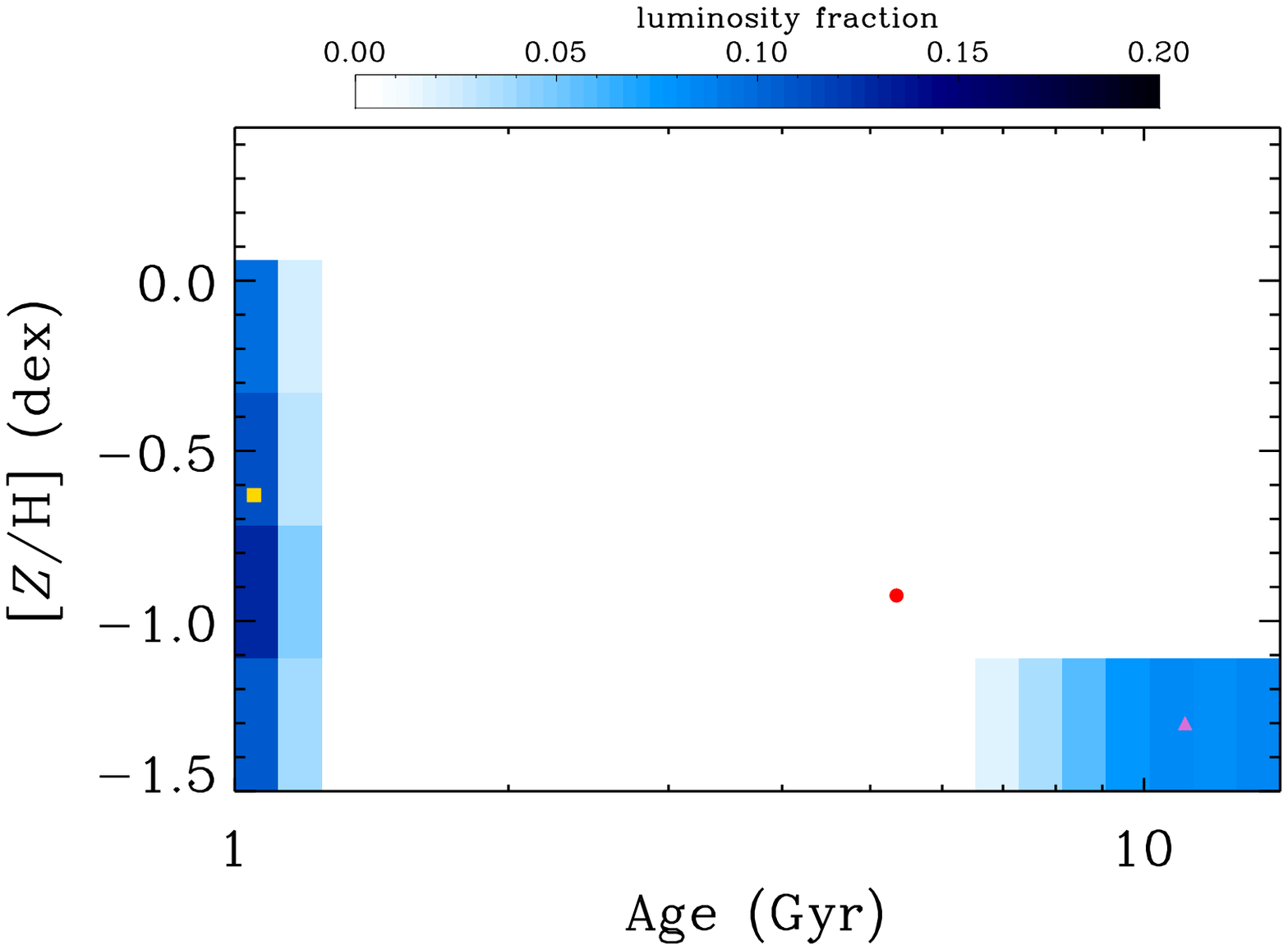}\\
 \centering{ESO-LV~4000370\\}
 \includegraphics[angle=0.0,width=0.431\textwidth,height=0.24\textheight]{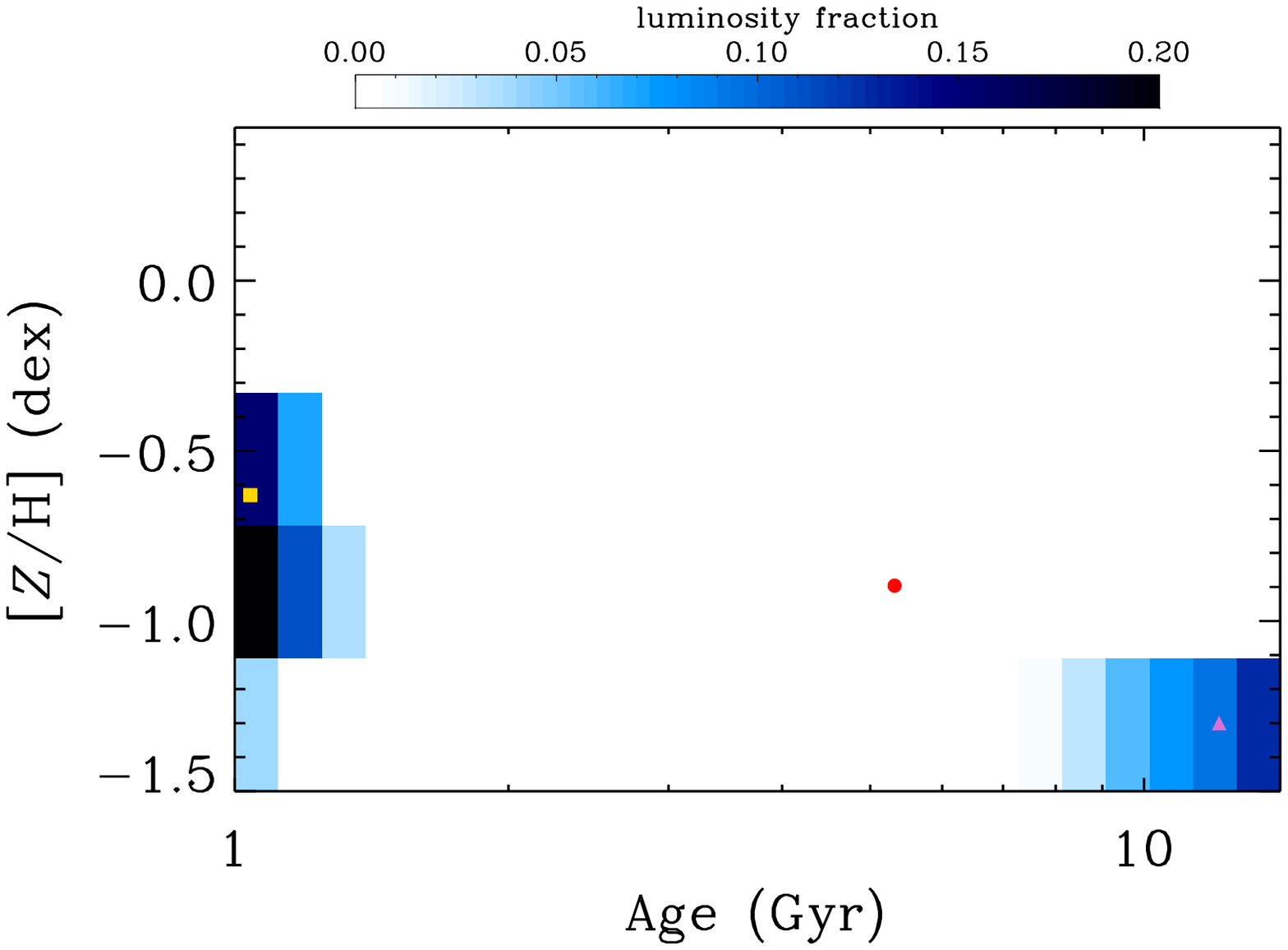}
 \includegraphics[angle=0.0,width=0.431\textwidth,height=0.24\textheight]{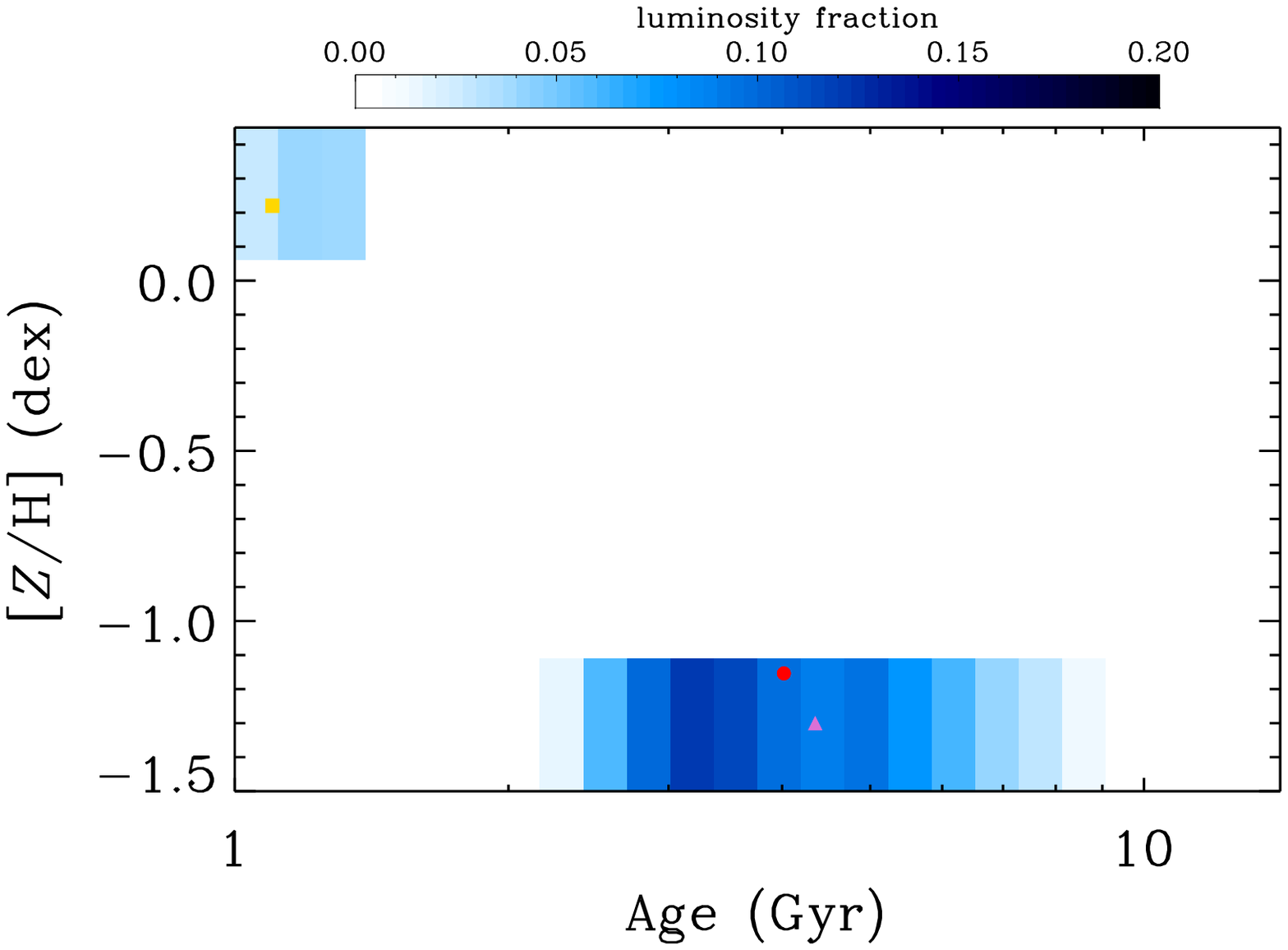}\\
 \caption{Age and metallicty distribution obtained from the
   best-fitting synthetic population models of the spectra of the
   sample galaxies obtained at \Rnc\ (left-hand panels) and
   \Rlast\ (right-hand panels). The colour scale gives the luminosity
   fraction in each bin of age and metallicity. The red circle marks
   the luminosity-weighted averages of age (\agelum) and metallicity
   (\metlum). The yellow square and purple triangle refer to the
   luminosity-weighted averages of age and metallicity for the old
   ($\rm Age> 4$ Gyr) and young ($\rm Age\leq 4$ Gyr) components of
   the stellar population, respectively.}
 \label{fig:age_metallicity}
\end{figure*}

\addtocounter{figure}{-1}
\begin{figure*}
 \centering{ESO-LV~4500200\\}
 \includegraphics[angle=0.0,width=0.431\textwidth,height=0.24\textheight]{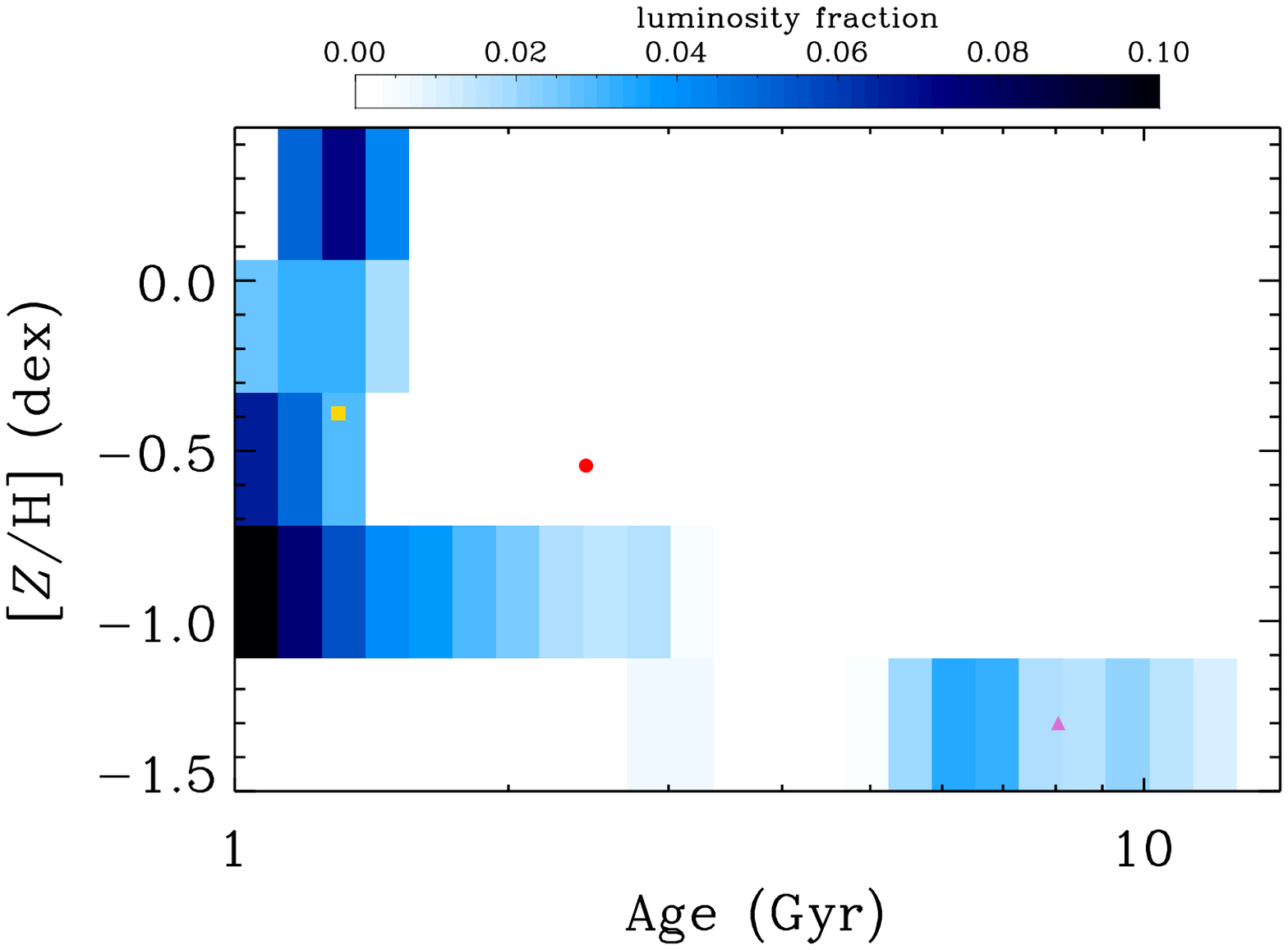}
 \includegraphics[angle=0.0,width=0.431\textwidth,height=0.24\textheight]{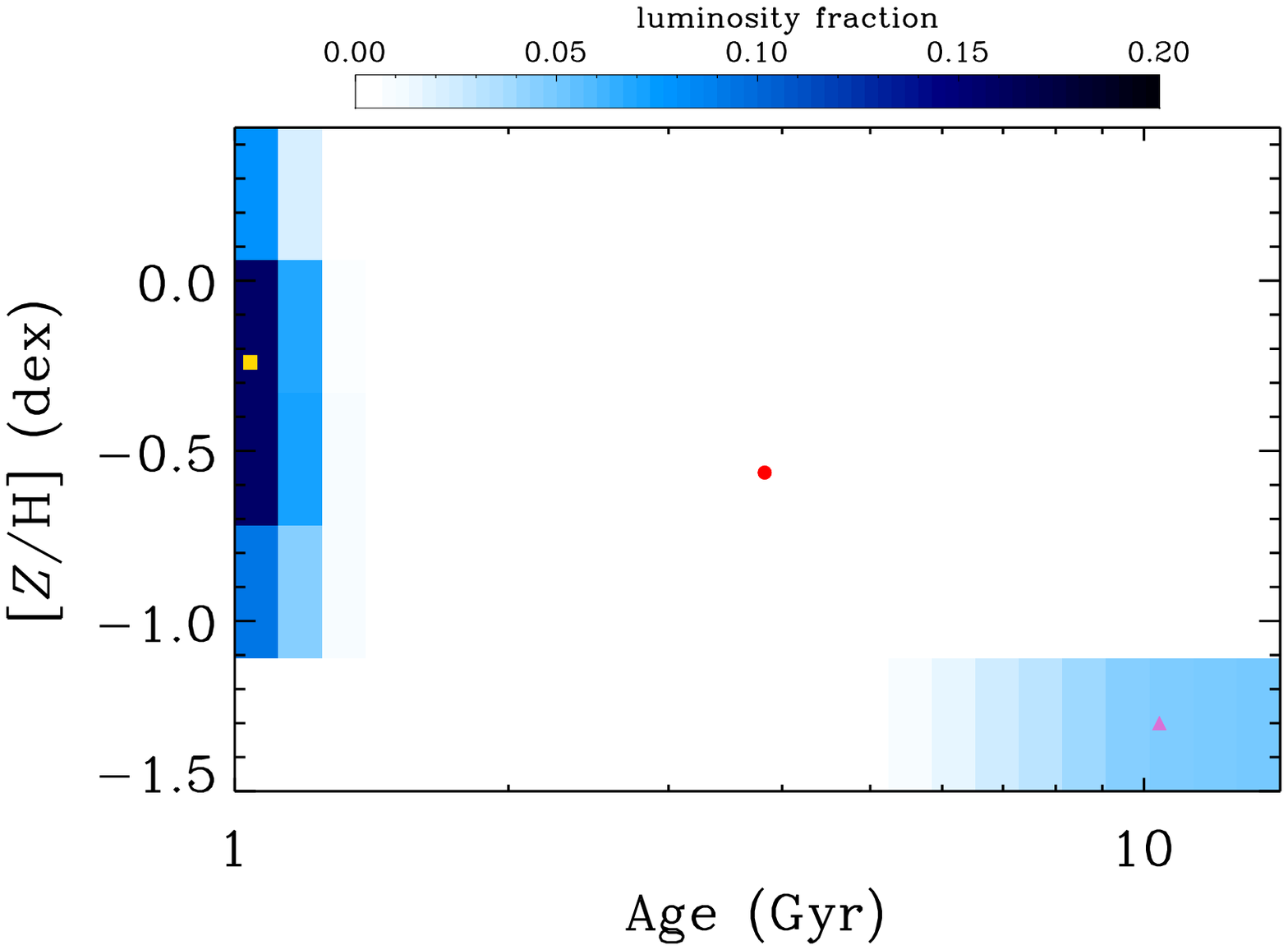}\\
 \centering{ESO-LV~5140100\\}
 \includegraphics[angle=0.0,width=0.431\textwidth,height=0.24\textheight]{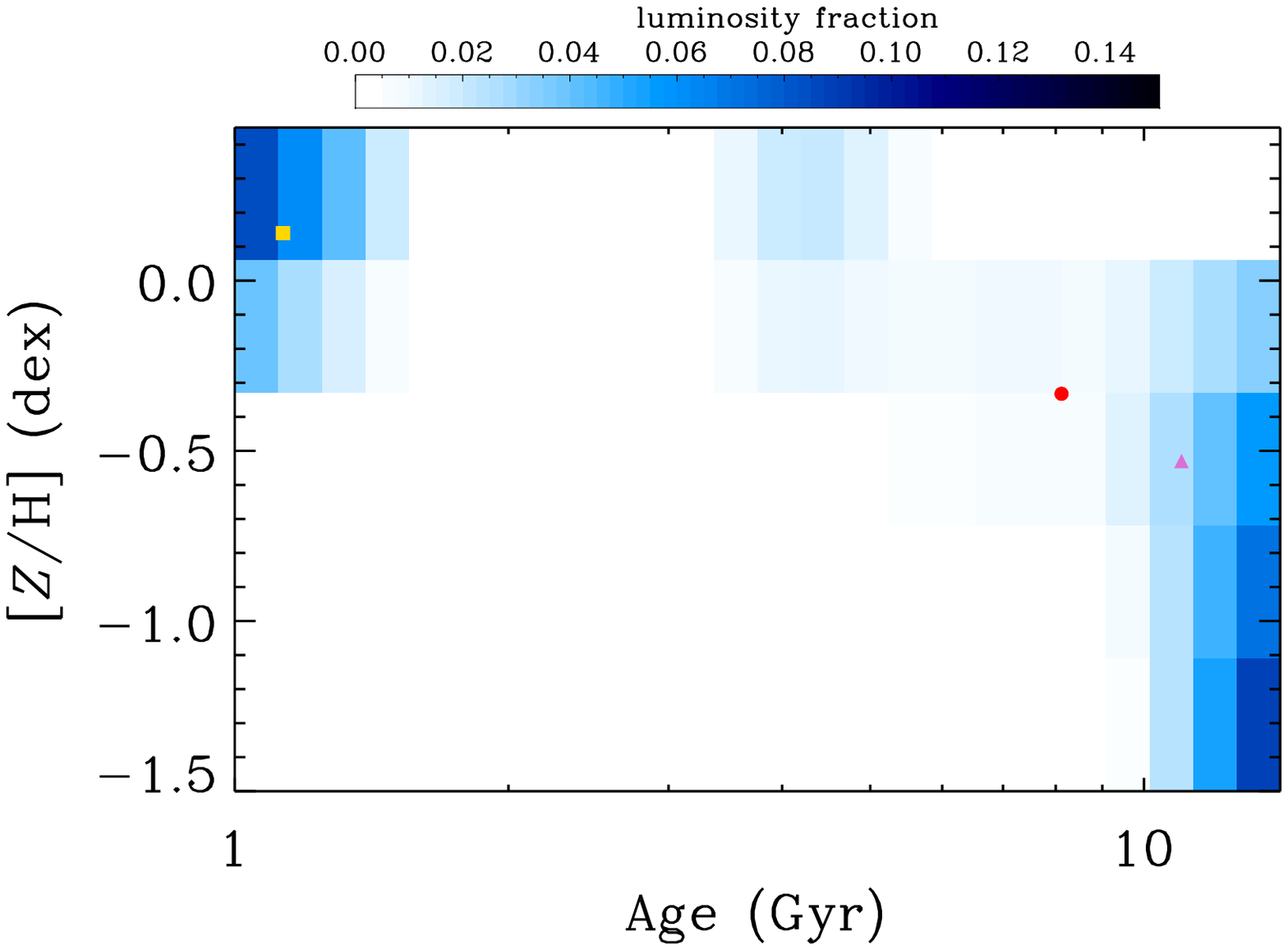}
 \includegraphics[angle=0.0,width=0.431\textwidth,height=0.24\textheight]{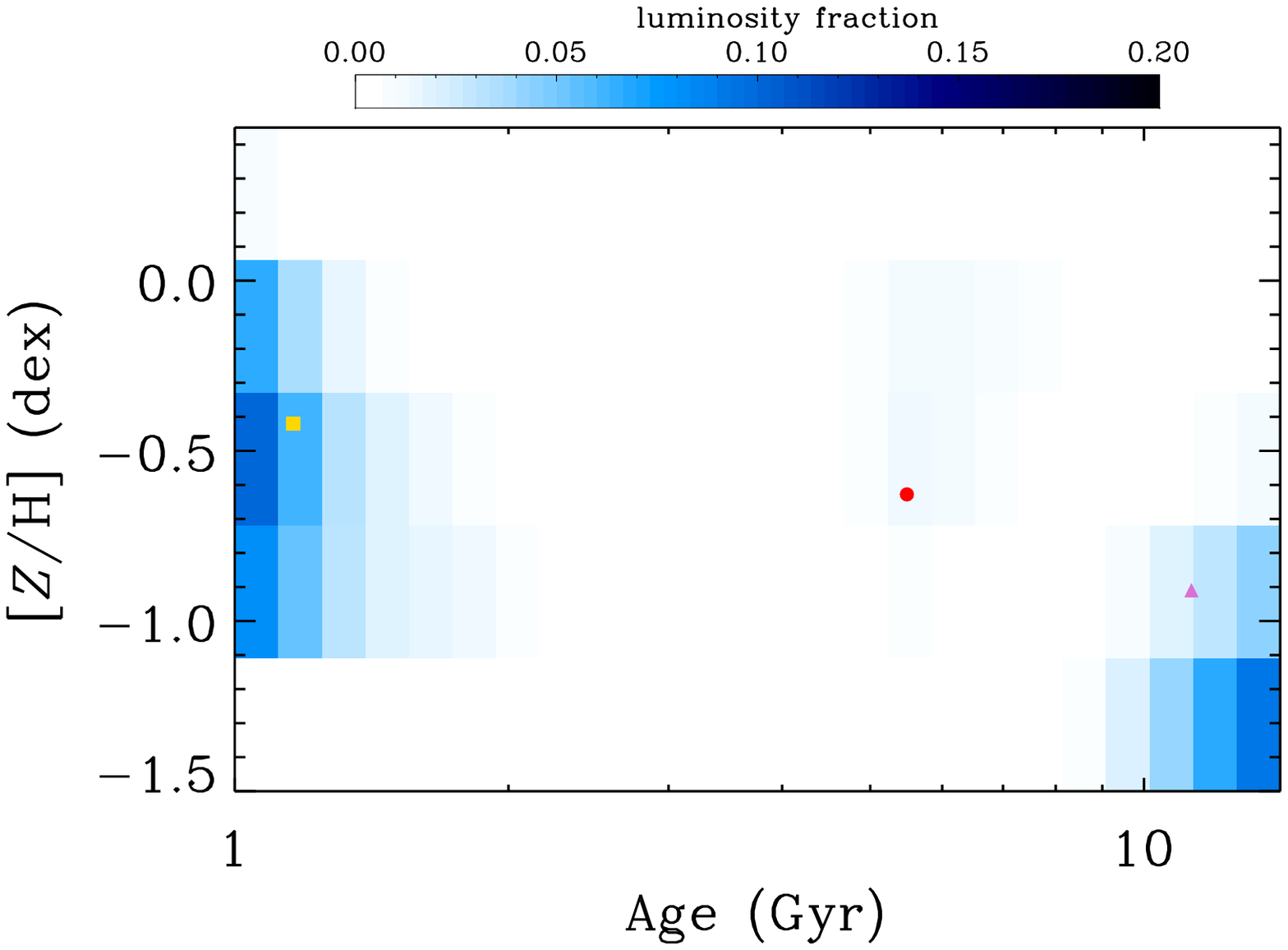}\\
 \centering{ESO-LV~5480440\\}
 \includegraphics[angle=0.0,width=0.431\textwidth,height=0.24\textheight]{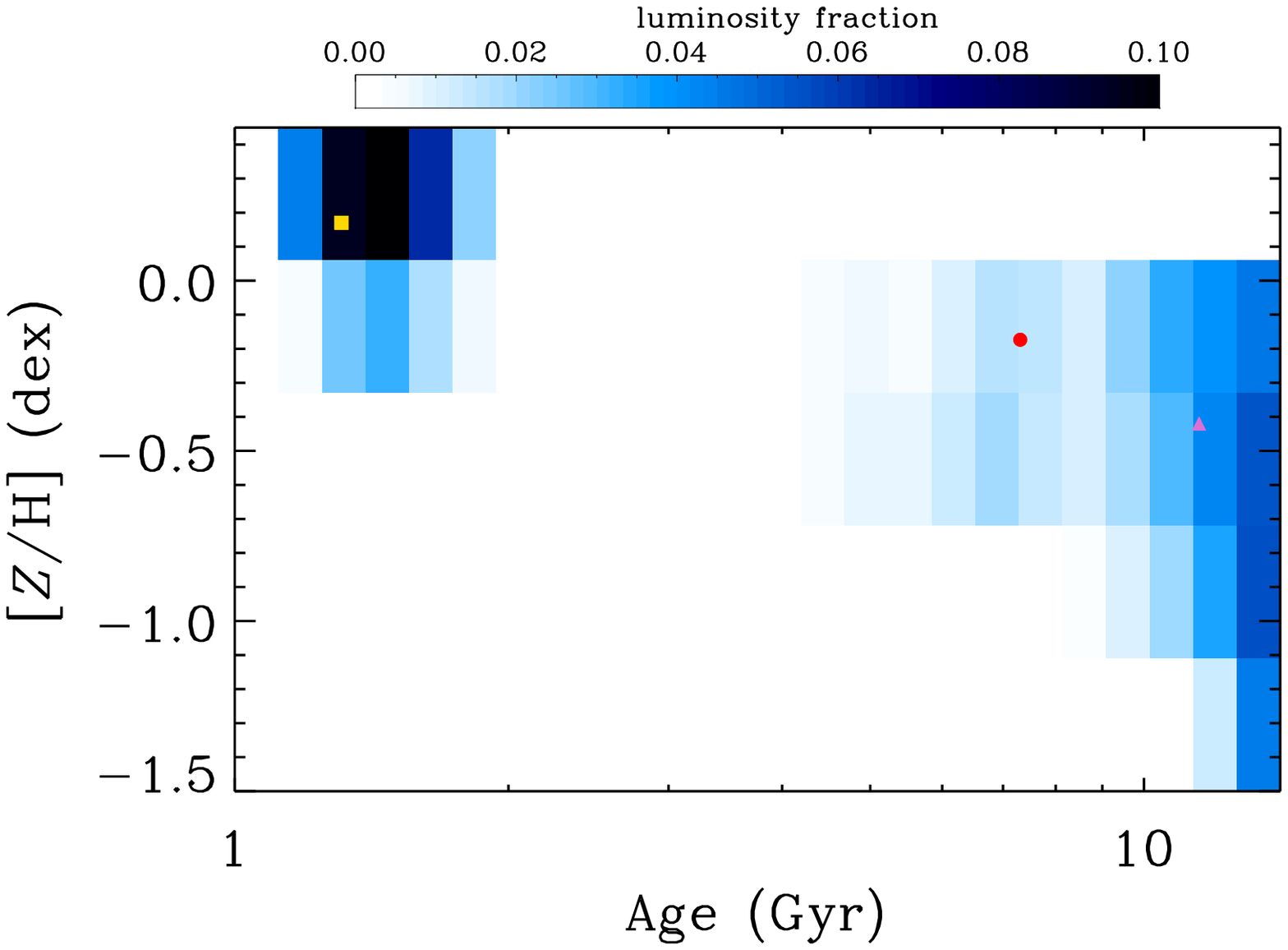}
 \includegraphics[angle=0.0,width=0.431\textwidth,height=0.24\textheight]{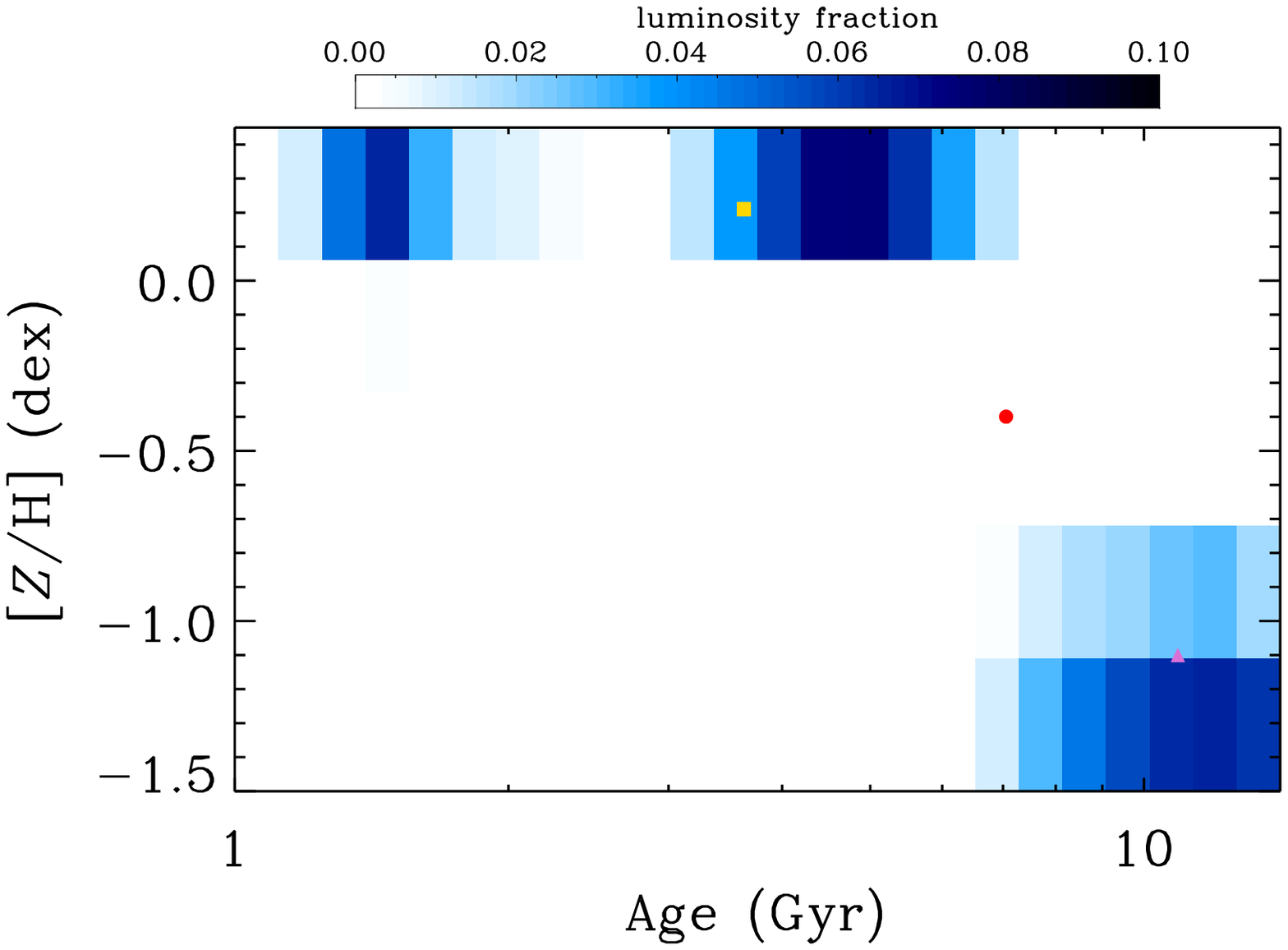}\\
 \caption{-- {\em continued}}
\end{figure*}

\addtocounter{figure}{-1}
 \begin{figure*}               
 \centering{IC~1993\\}
 \includegraphics[angle=0.0,width=0.431\textwidth,height=0.24\textheight]{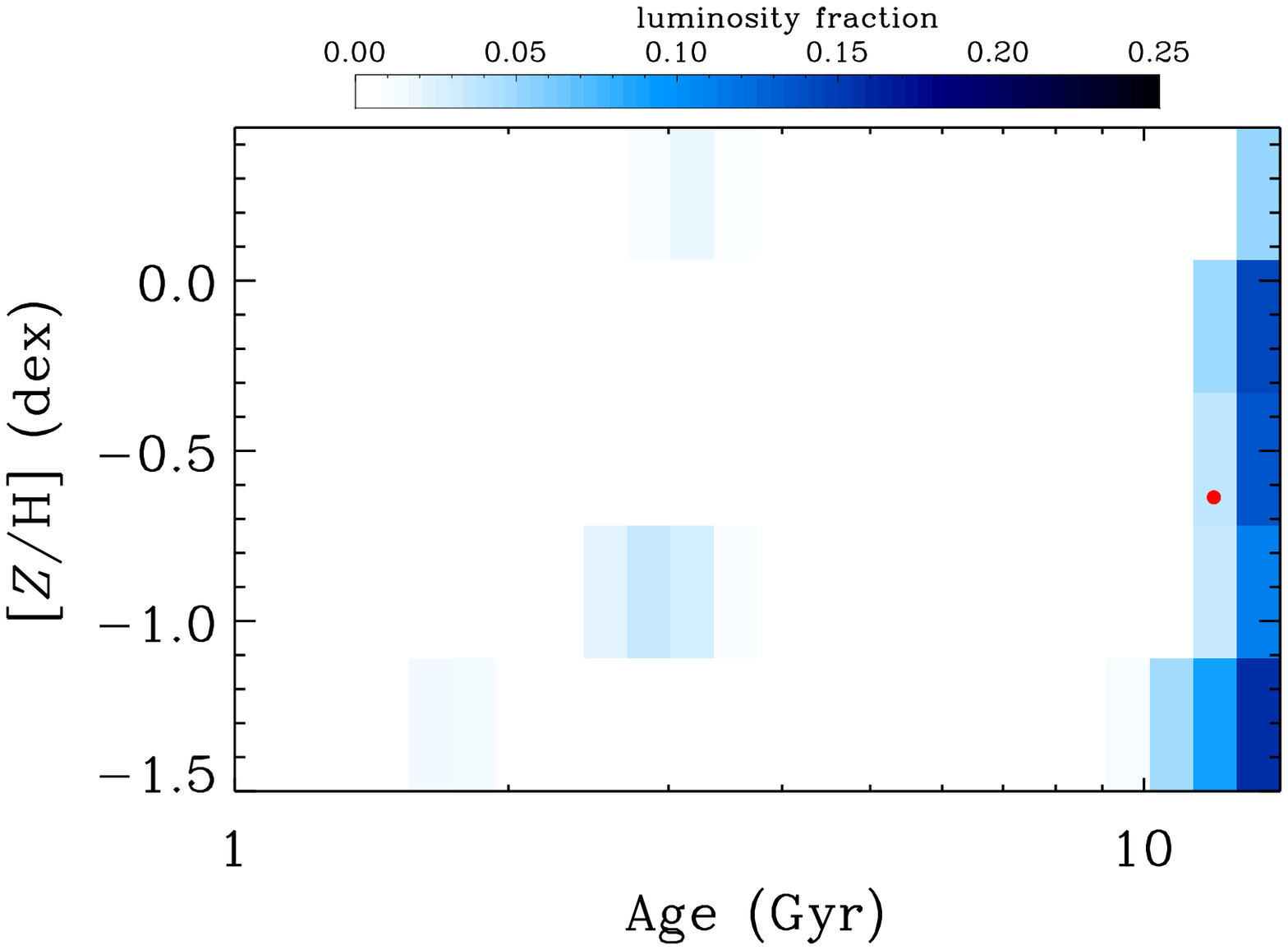}
 \includegraphics[angle=0.0,width=0.431\textwidth,height=0.24\textheight]{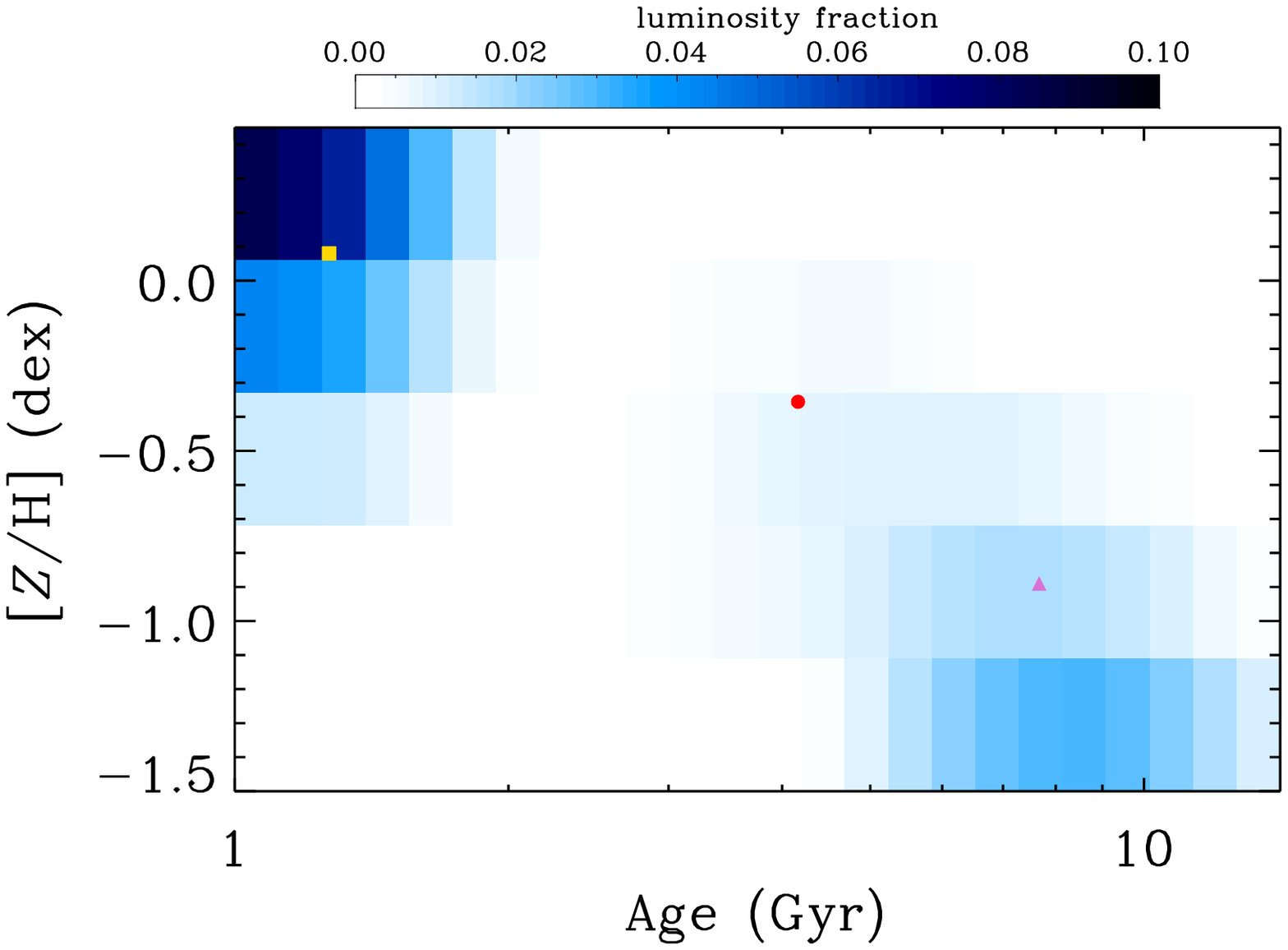}\\
 \centering{NGC~1366\\}
 \includegraphics[angle=0.0,width=0.431\textwidth,height=0.24\textheight]{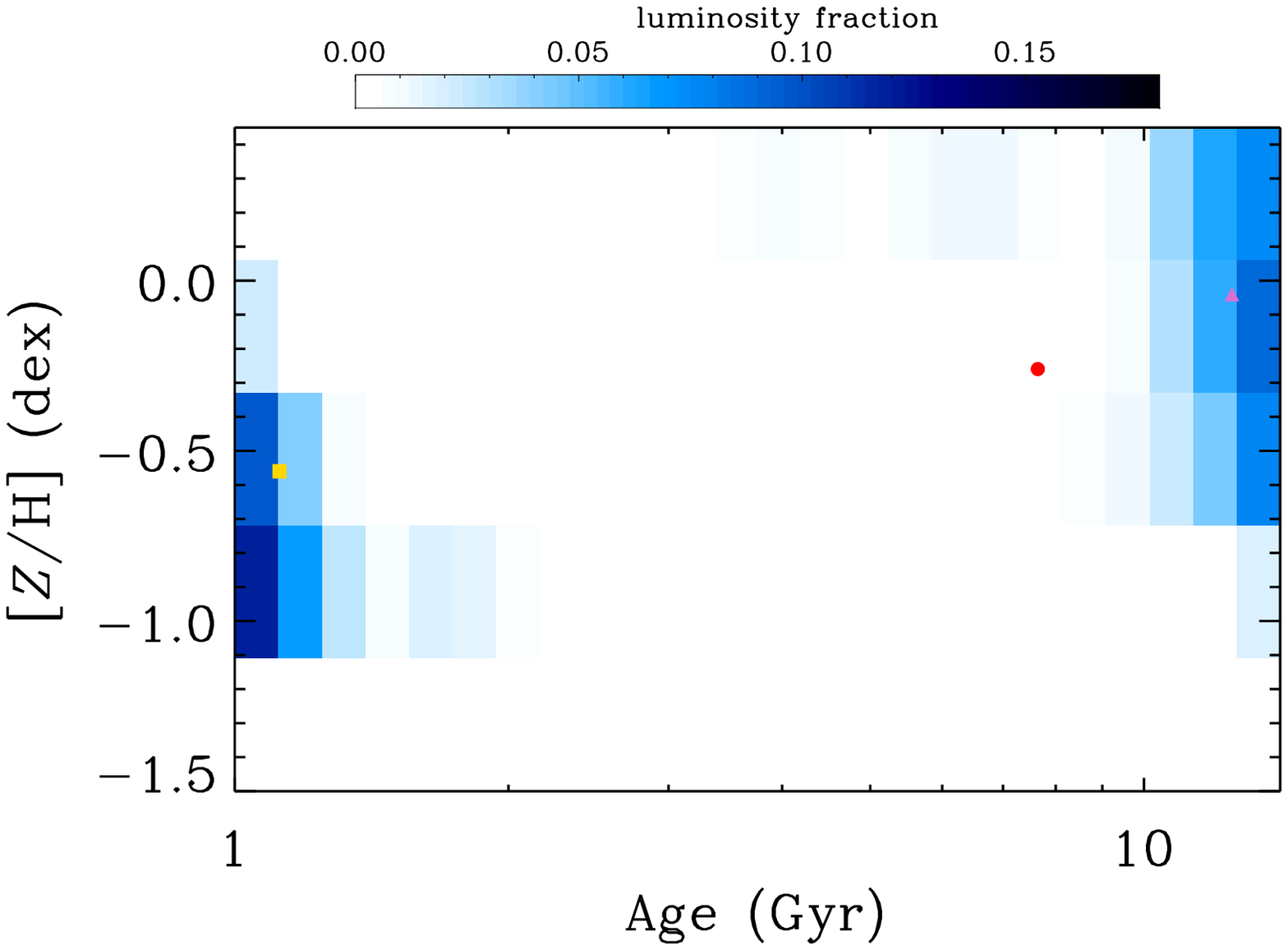}
 \includegraphics[angle=0.0,width=0.431\textwidth,height=0.24\textheight]{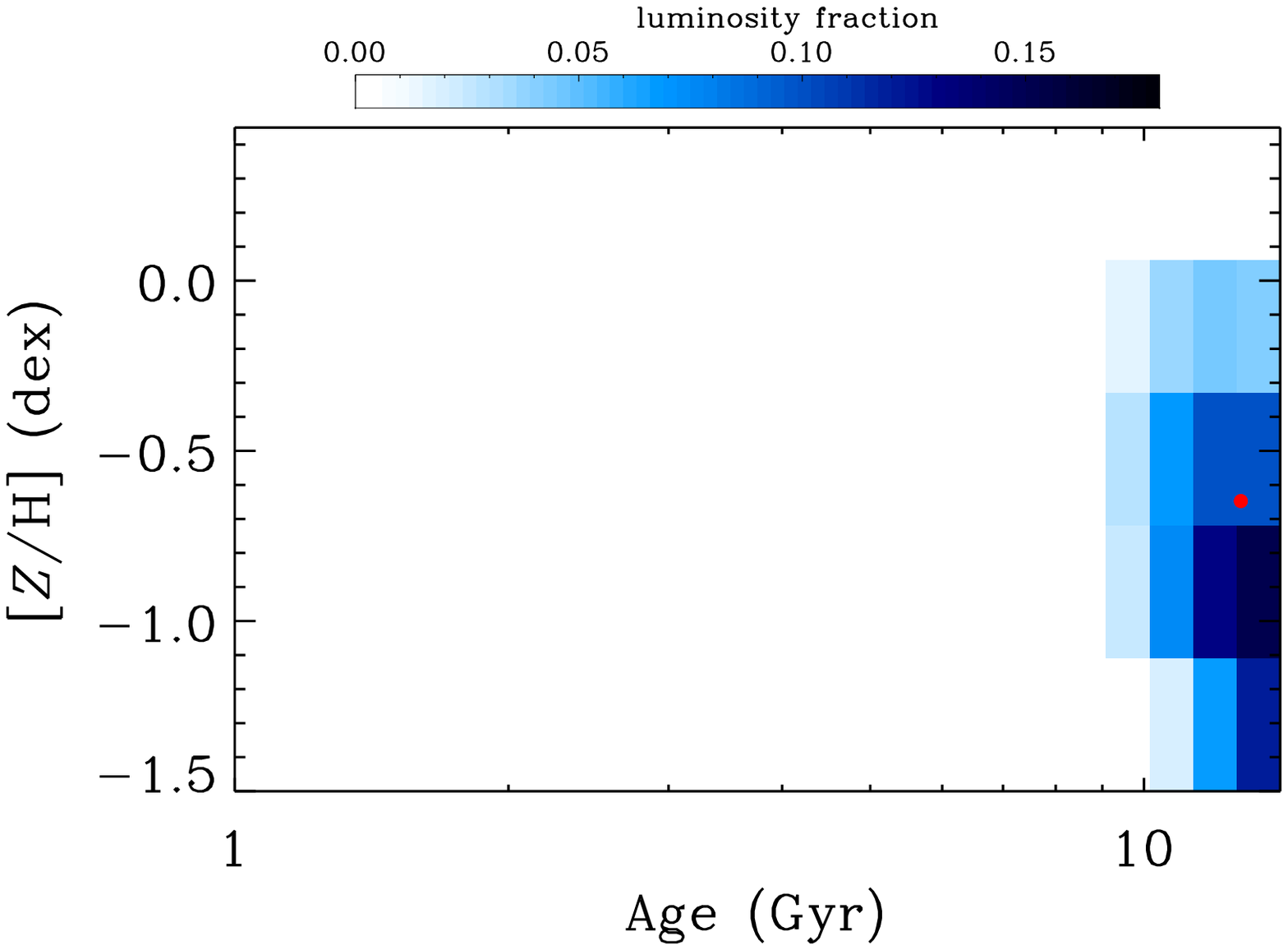}\\
 \centering{NGC~7643\\}
 \includegraphics[angle=0.0,width=0.431\textwidth,height=0.24\textheight]{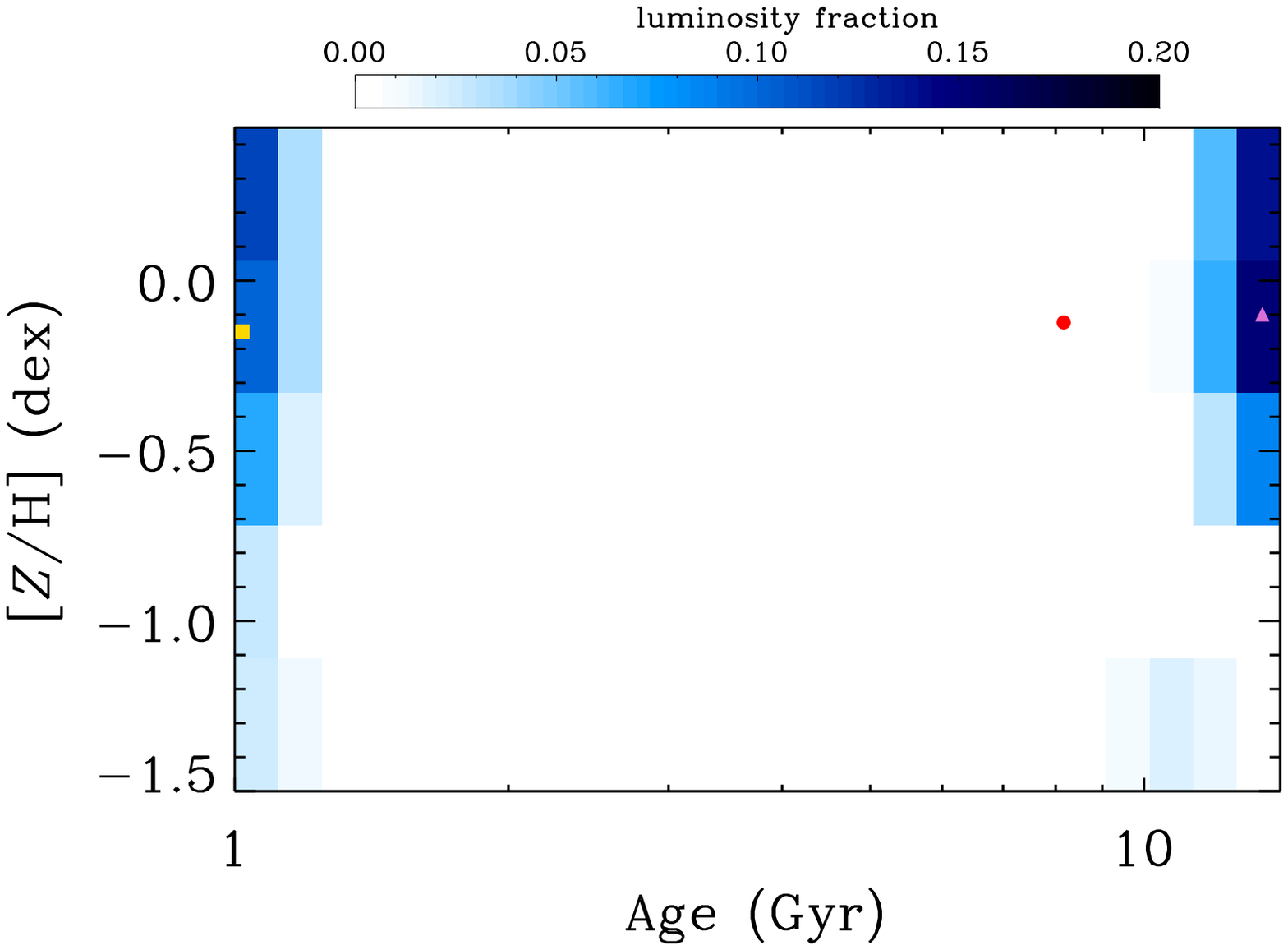}
 \includegraphics[angle=0.0,width=0.431\textwidth,height=0.24\textheight]{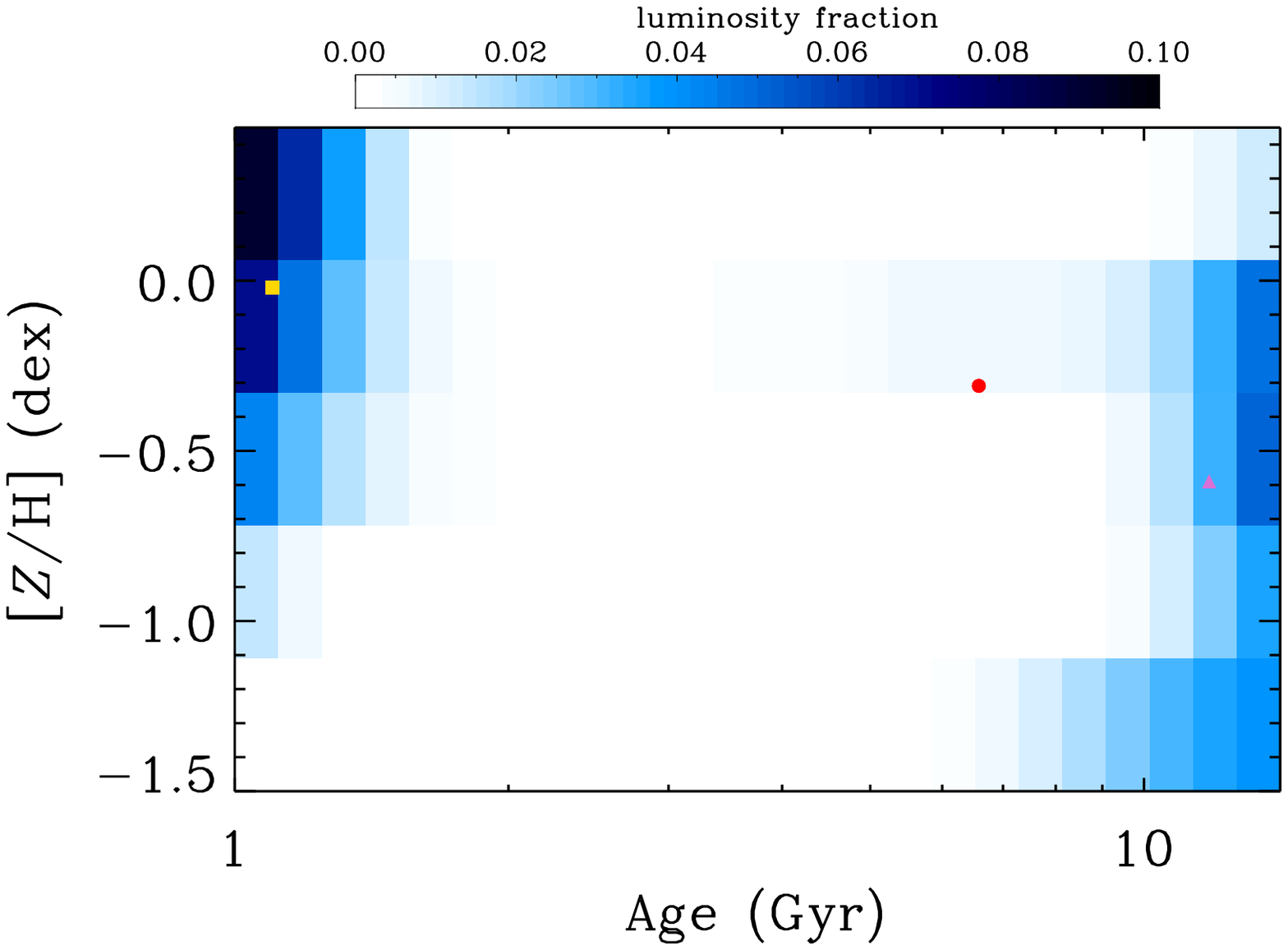}\\
 \caption{-- {\em continued}}
\end{figure*}

\addtocounter{figure}{-1}
 \begin{figure*}               
 \centering{PGC~37759\\}
 \includegraphics[angle=0.0,width=0.431\textwidth,height=0.24\textheight]{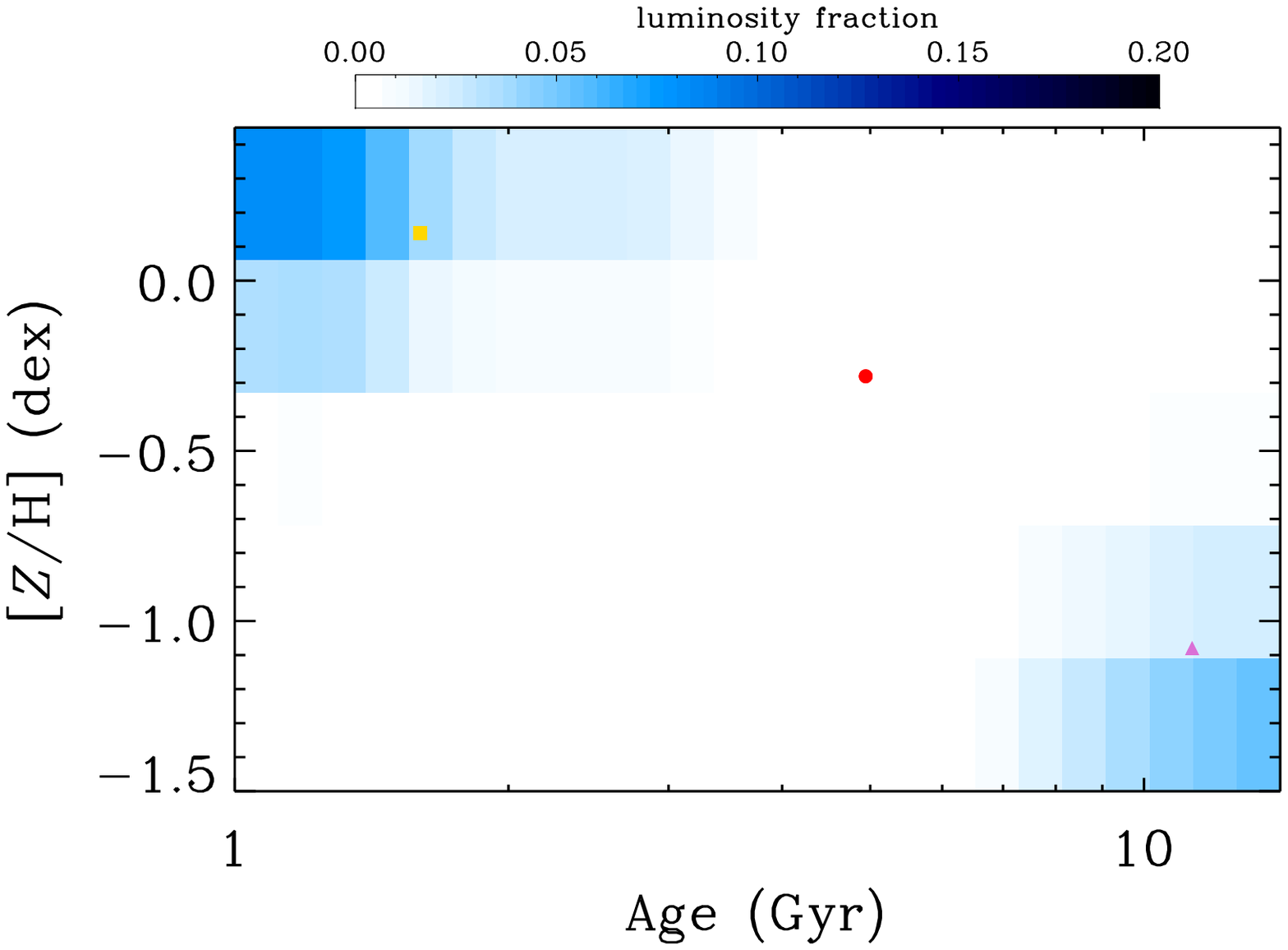}
 \includegraphics[angle=0.0,width=0.431\textwidth,height=0.24\textheight]{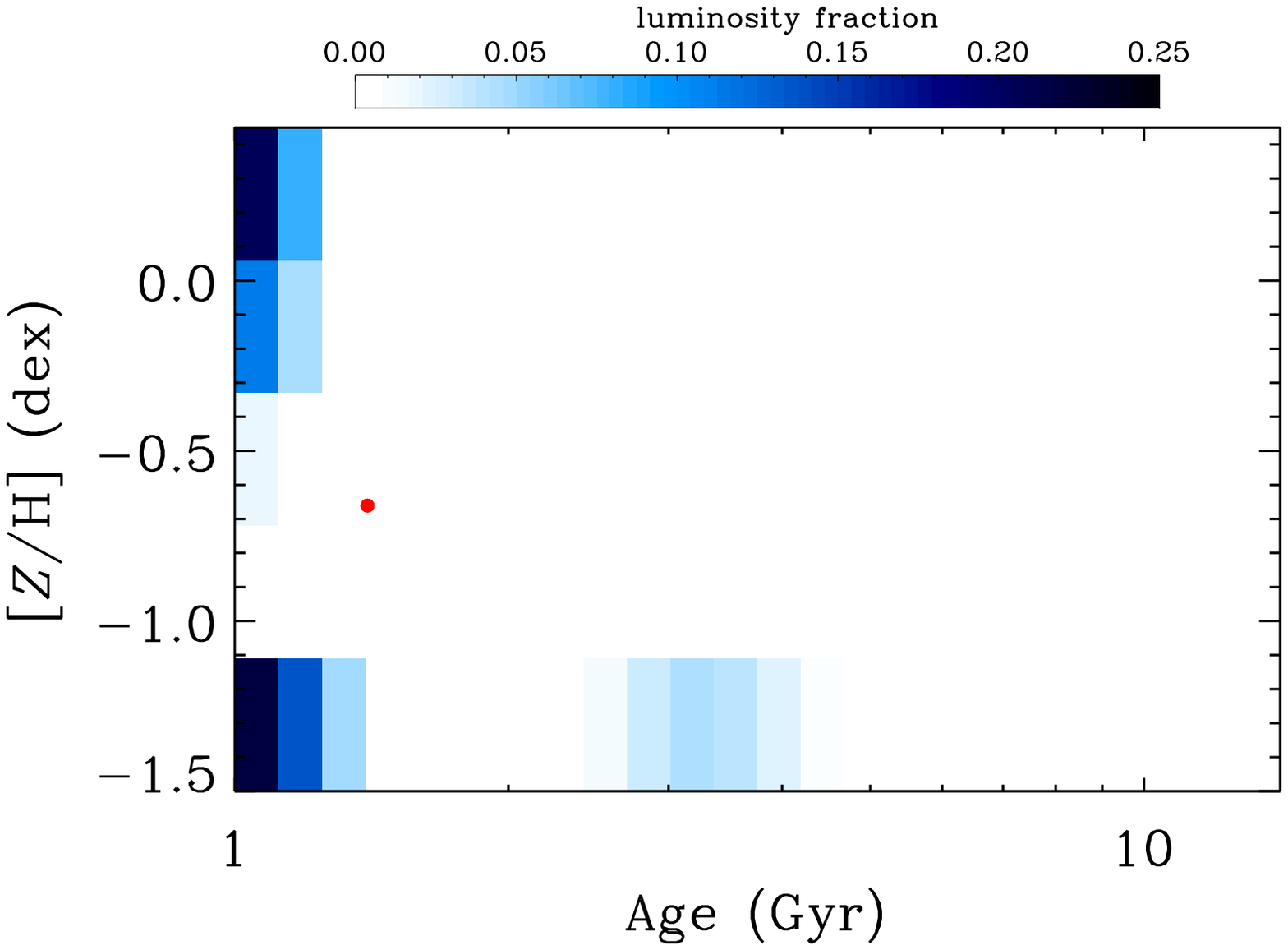}\\
 \caption{-- {\em continued}}
\end{figure*}

\subsubsection{Luminosity-weighted age and metallicity}
\label{sec:1burst}

\begin{figure}
 \includegraphics[angle=90.0,width=0.5\textwidth]{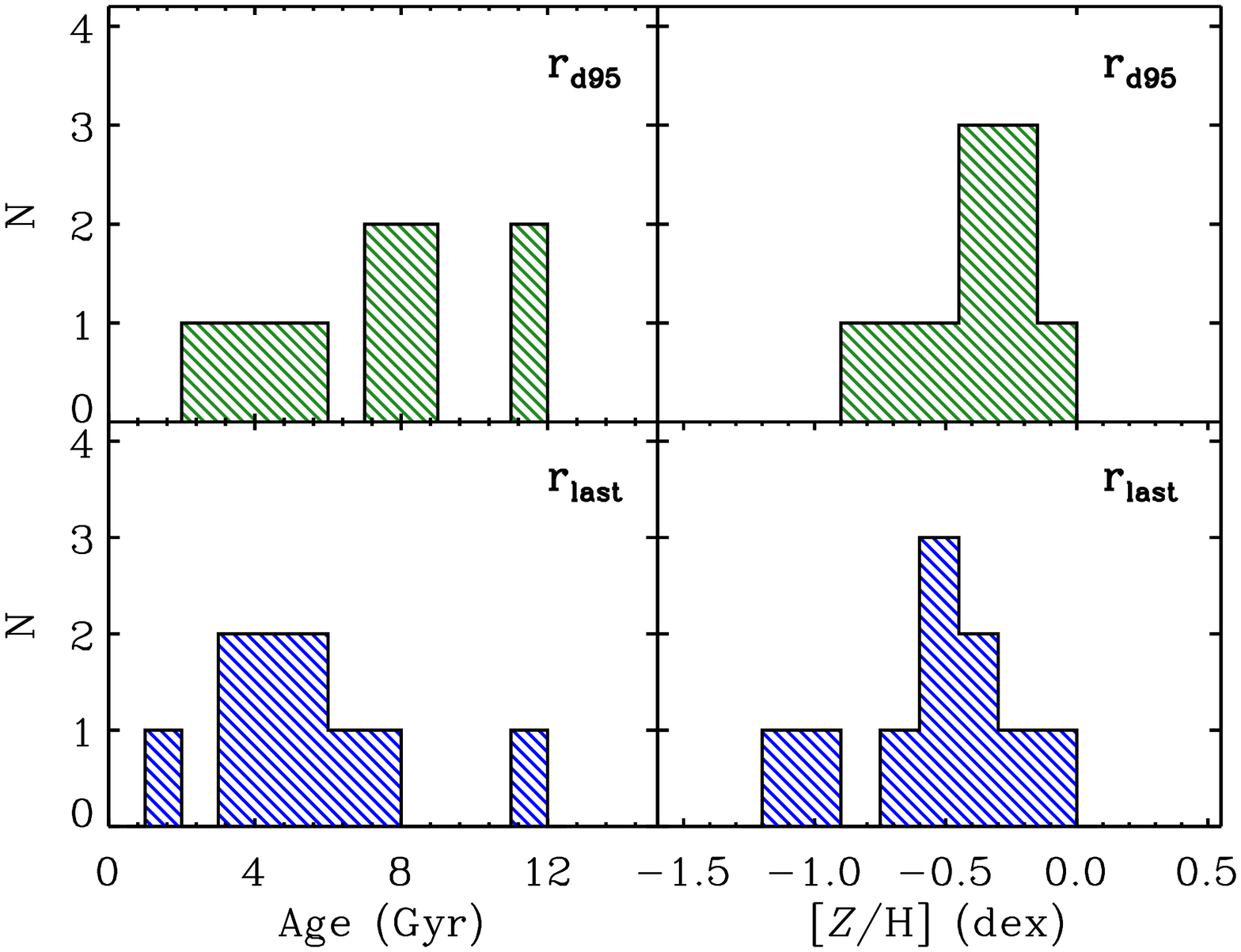}
 \caption{Distribution of the luminosity-weighted age (left-hand
   panels) and metallicity (right-hand panels) calculated at
   \Rnc\ (green histograms, upper panels) and \Rlast\ (blue
   histograms, lower panels) for the stellar populations in the discs
   of the sample galaxies.}
 \label{fig:pops_histo_1burst}
\end{figure}

For all the sample galaxies we calculated the luminosity-weighted age
\agelum\ and metallicity \metlum\ of the stellar population of their
discs at \Rnc\ and \Rlast . The uncertainties for ages and
metallicities were estimated by Monte Carlo simulations. For each
galaxy, we built 100 simulated spectra by adding to the best-fitting
synthetic population model a noise spectrum with the same standard
deviation of the difference between the observed and model
spectrum. The simulated spectra were measured as if they were
real. The standard deviations of the distributions of the simulated
ages and metallicities were adopted as errors on the measured age and
metallicity, respectively. The values are listed in Table
\ref{tab:results_1burst} and their number distributions are plotted in
Fig.~\ref{fig:pops_histo_1burst}.

No significant difference was found between the distributions of the
luminosity-weighted ages at \Rnc\/ and \Rlast, both of them spanning a
large range ($1\,\leq\,$\agelum$\,\leq12\,$ Gyr).  On the contrary,
the distribution of the luminosity-weighted metallicity at \Rlast\ is
characterised by lower values and it peaks at \metlum$\,\simeq-0.5$
dex, while the luminosity-weighted metallicities at \Rnc\/ are
slightly shifted to higher values with a peak at \metlum$\,\simeq-0.4$
dex.

\begin{figure}
 \includegraphics[angle=90.0,width=0.5\textwidth]{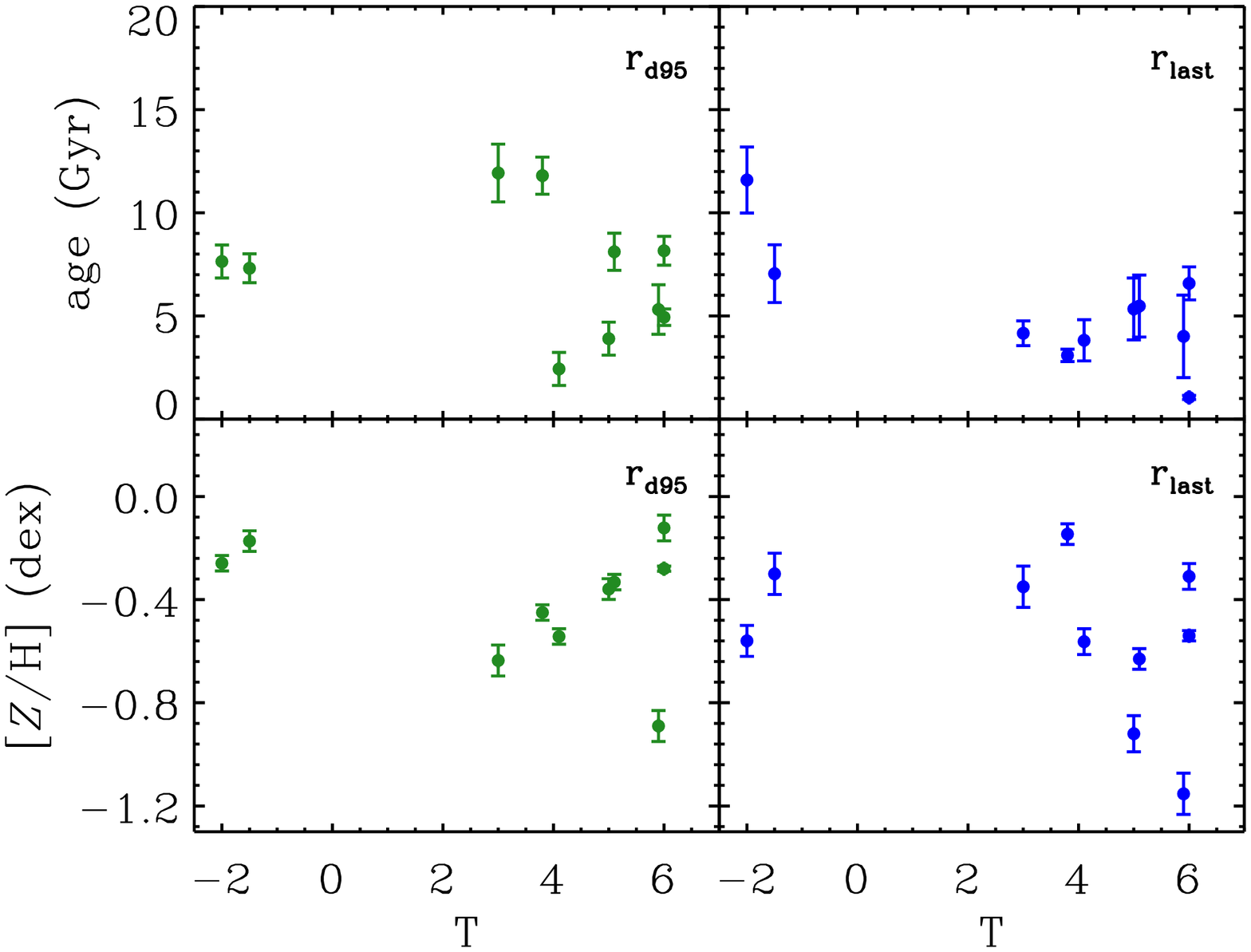}
 \caption{Luminosity-weighted age (upper panels) and
   luminosity-weighted metallicity (lower panels) measured at
   \Rnc\ (green circles, left-hand panels) and \Rlast\ (blue circles,
   right-hand panels) for the stellar populations in the discs of the
   sample galaxies as a function of their morphological type. }
 \label{fig:pops_type_1burst}
\end{figure}

The bulge-to-disc ratio is one of the key ingredients for the
morphological classification of galaxies. Our sample covers the Hubble
sequence from S0s to Sc spirals, but there is no evidence for a clear
cut correlation between the galaxy type and stellar population
properties of the disc-dominated region
(Fig.~\ref{fig:pops_type_1burst}).

\begin{table*}
\caption{Age and metallicity of the stellar populations in the
  disc-dominated region of the sample galaxies. The columns show the
  following: 1, galaxy name; 2, radius where the disc contributes more
  than $95\%$ of the galaxy surface brightness; 3, farthest radius
  where the stellar population properties are measured; 4, radial
  range \Rlast$-$\Rnc\ normalized to the disc scalelength $h$, where
  the luminosity-weighted age and metallicty gradients are calculated;
  5, luminosity-weighted age at \Rnc ; 6, luminosity-weighted
  metallicity at \Rnc ; 7, luminosity-weighted age at \Rlast ; 8,
  luminosity-weighted metallicity at \Rlast ; 9, gradient of the
  luminosity-weighted age in the disc; 10, gradient of the
  luminosity-weighted metallicity in the disc. }
\label{tab:results_1burst}
\begin{center}
\begin{small}
\begin{tabular}{l ccc cccc rr}
\hline
\noalign{\smallskip}
\multicolumn{1}{c}{} & 
\multicolumn{1}{c}{} &
\multicolumn{1}{c}{} &
\multicolumn{1}{c}{} &
\multicolumn{2}{c}{$r_{\rm{ d95}}$} & 
\multicolumn{2}{c}{$r_{\rm{ last}}$} &
\multicolumn{1}{c}{} &
\multicolumn{1}{c}{} \\
\multicolumn{1}{c}{Galaxy} & 
\multicolumn{1}{c}{\Rnc} &
\multicolumn{1}{c}{\Rlast} &
\multicolumn{1}{c}{$\Delta r/h$} &
\multicolumn{1}{c}{\agelum} & 
\multicolumn{1}{c}{\metlum} &
\multicolumn{1}{c}{\agelum} &
\multicolumn{1}{c}{\metlum} &
\multicolumn{1}{c}{$\Delta$\agelum} &
\multicolumn{1}{c}{$\Delta$\metlum} \\
\multicolumn{1}{c}{} &
\multicolumn{1}{c}{(arcsec)} &
\multicolumn{1}{c}{(arcsec)} &
\multicolumn{1}{c}{} &
\multicolumn{1}{c}{(Gyr)} & 
\multicolumn{1}{c}{(dex)}&
\multicolumn{1}{c}{(Gyr)} & 
\multicolumn{1}{c}{(dex)}&
\multicolumn{1}{c}{(Gyr)} & 
\multicolumn{1}{c}{(dex)} \\
\multicolumn{1}{c}{(1)} &
\multicolumn{1}{c}{(2)} &
\multicolumn{1}{c}{(3)} &
\multicolumn{1}{c}{(4)} &
\multicolumn{1}{c}{(5)} &
\multicolumn{1}{c}{(6)} &
\multicolumn{1}{c}{(7)} &
\multicolumn{1}{c}{(8)} &
\multicolumn{1}{c}{(9)} &
\multicolumn{1}{c}{(10)} \\ 
\noalign{\smallskip}
\hline		    
\noalign{\smallskip}        	                		 
ESO-LV~1890070 & $12.2$ & $59.7$ &1.52 & $11.7\pm0.9$ & $-0.45\pm0.03$ & $ 3.1\pm0.3$ & $-0.15\pm0.04$ & $-5.66\pm0.78$ & $ 0.20\pm0.04$ \\ 
ESO-LV~2060140 &  $9.1$ & $37.5$ &1.57 & $ 3.9\pm0.8$ & $-0.36\pm0.04$ & $ 5.3\pm1.5$ & $-0.92\pm0.07$ & $ 0.92\pm1.46$ & $-0.36\pm0.07$ \\ 
ESO-LV~4000370 &  $5.7$ & $23.7$ &0.80 & $ 5.3\pm1.2$ & $-0.89\pm0.06$ & $ 4.0\pm2.0$ & $-1.15\pm0.08$ & $-1.61\pm3.97$ & $-0.33\pm0.17$ \\ 
ESO-LV~4500200 & $17.1$ & $54.0$ &2.13 & $ 2.4\pm0.8$ & $-0.54\pm0.03$ & $ 3.8\pm1.0$ & $-0.56\pm0.05$ & $ 0.65\pm0.84$ & $-0.01\pm0.03$ \\ 
ESO-LV~5140100 &  $6.1$ & $58.2$ &1.91 & $ 8.1\pm0.9$ & $-0.33\pm0.03$ & $ 5.5\pm1.5$ & $-0.63\pm0.04$ & $-1.38\pm1.25$ & $-0.16\pm0.03$ \\ 
ESO-LV~5480440 &  $5.0$ & $21.4$ &1.63 & $ 7.3\pm0.7$ & $-0.17\pm0.04$ & $ 7.1\pm1.4$ & $-0.30\pm0.08$ & $-0.16\pm1.29$ & $-0.01\pm0.07$ \\ 
IC~1993        &  $7.2$ & $27.5$ &0.93 & $11.9\pm1.4$ & $-0.64\pm0.06$ & $ 4.2\pm0.6$ & $-0.35\pm0.08$ & $-8.39\pm2.15$ & $ 0.31\pm0.15$ \\ 
NGC~1366       & $12.5$ & $26.2$ &1.06 & $ 7.6\pm0.8$ & $-0.26\pm0.03$ & $11.6\pm1.6$ & $-0.56\pm0.06$ & $ 3.72\pm2.26$ & $-0.28\pm0.08$ \\ 
NGC~7643       &  $4.5$ & $15.8$ &1.02 & $ 8.2\pm0.7$ & $-0.12\pm0.05$ & $ 6.6\pm0.8$ & $-0.31\pm0.05$ & $-1.55\pm1.46$ & $-0.18\pm0.09$ \\ 
PGC~37759      &  $1.5$ & $11.1$ &1.34 & $ 4.9\pm0.4$ & $-0.28\pm0.01$ & $ 1.1\pm0.1$ & $-0.54\pm0.02$ & $-2.91\pm0.37$ & $-0.19\pm0.02$ \\ 
\noalign{\smallskip}
\hline				    	    			 
\end{tabular}
\end{small}
\end{center}
\end{table*}

\begin{figure}
\centering
\includegraphics[angle=90,width=0.5\textwidth]{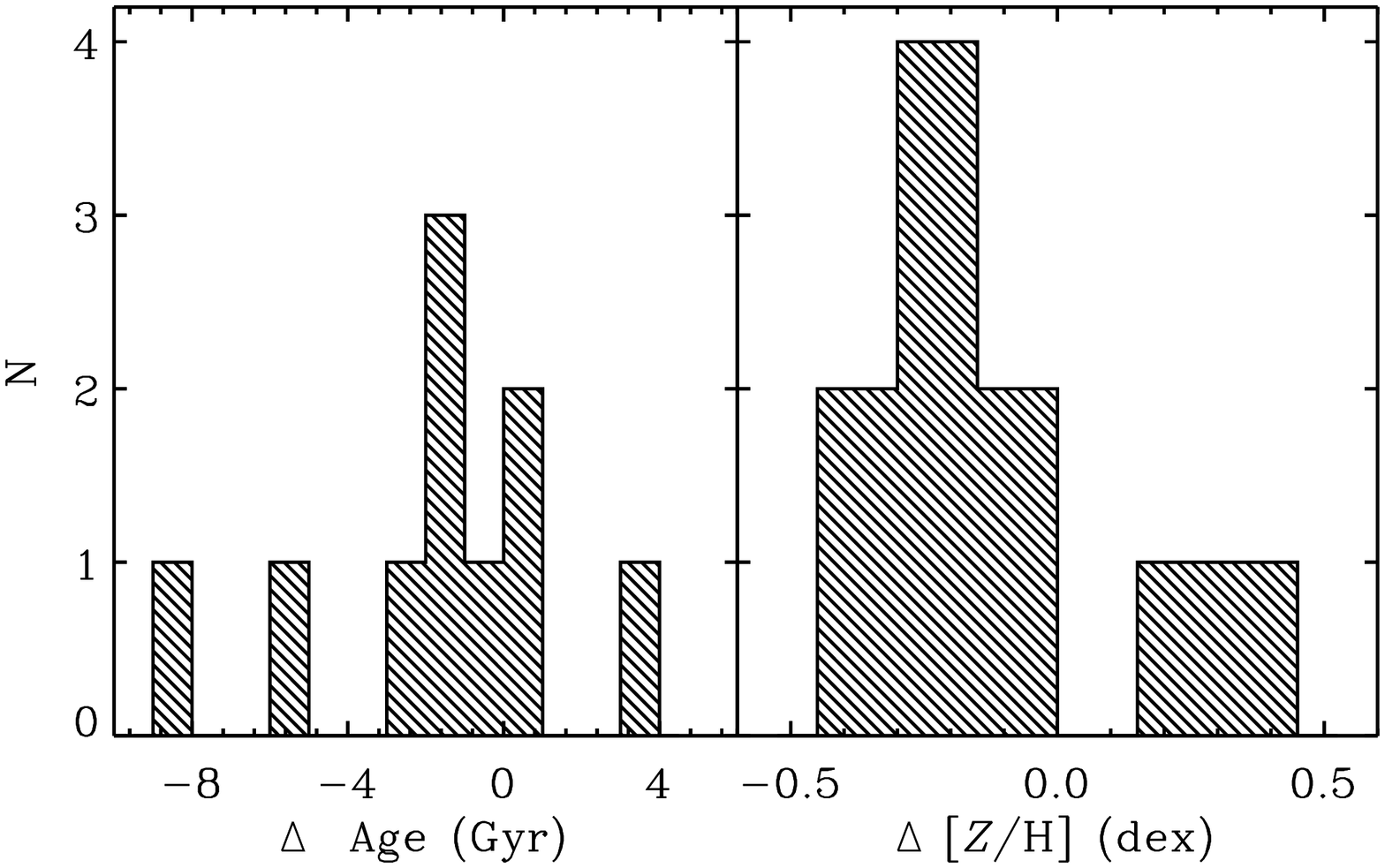}\\
 \caption{Distribution of the gradients of the luminosity-weighted age
   (left-hand panel) and metallicity (right-hand panel) of the stellar
   populations in the discs of the sample galaxies.}
\label{fig:grad_histo_1burst}
\end{figure}

We calculated the gradients of the luminosity-weighted age and
metallicity over the disc scalelength from the values measured at
\Rnc\ and \Rlast. The gradients and corresponding errors are given in
Table~\ref{tab:results_1burst} and their number distributions are
plotted in Fig.~\ref{fig:grad_histo_1burst}.

The age gradient is negligible within the errors in most of the discs.
ESO-LV~1890070, IC~1993, and PGC~377759 display a negative gradient,
whereas NGC~1366 has a remarkably strong positive gradient.
The metallicity gradient of all the discs is negative or null, except
ESO-LV~1890070 and IC~1993. This is expected if the disc components
assembled through an inside-out or outside-in process
\citep{brook2004, munoetal07}. 
The distribution of the age and metallicity gradients is consistent
with that of the unbarred disc galaxies studied by \citet{sancetal14}
once their gradients are rescaled to the disc scalelength.

\begin{figure}
 \includegraphics[angle=90,width=0.5\textwidth]{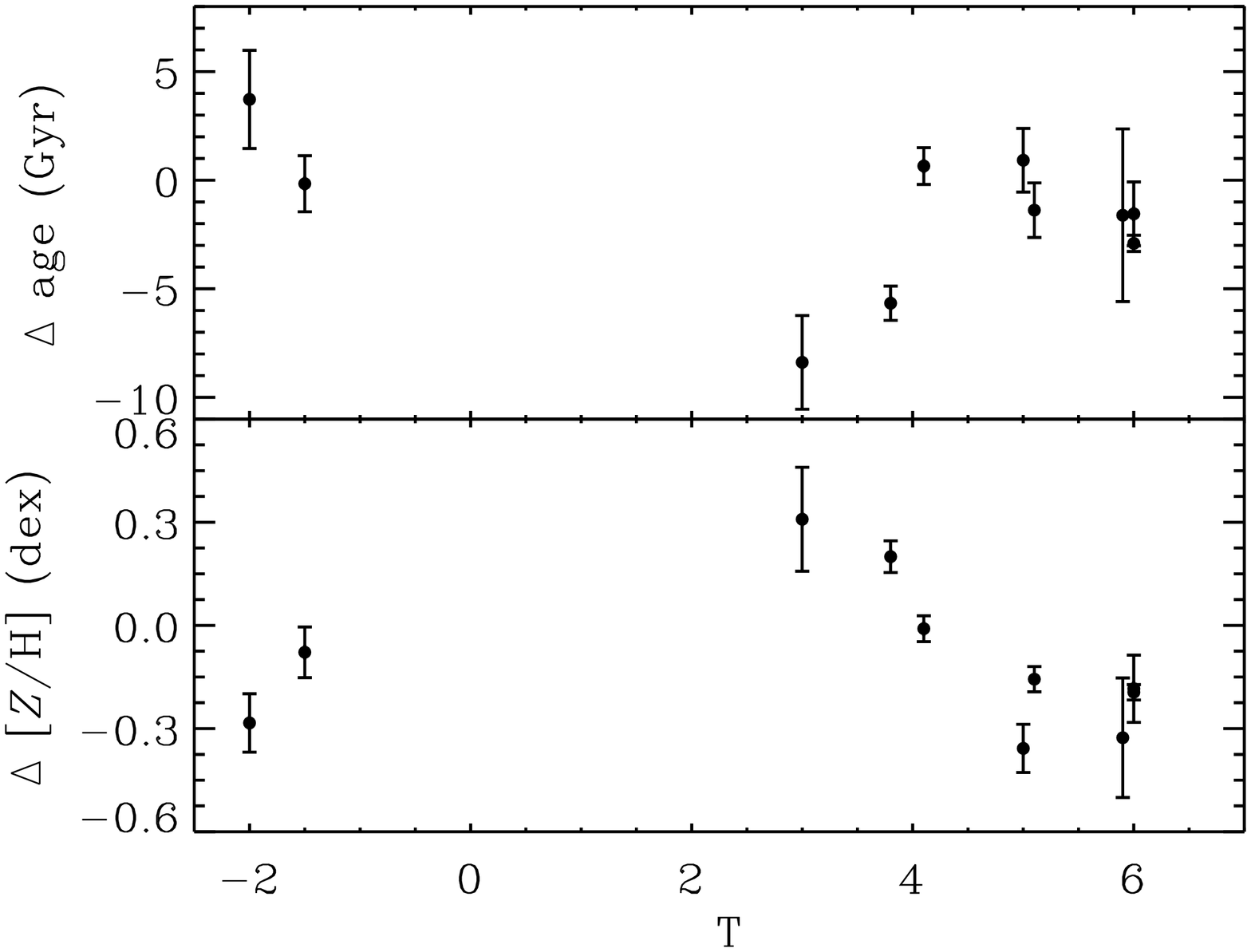}
 \caption{Gradients of the luminosity-weighted age (upper panel) and
   metallicity (lower panel) of the stellar populations in the discs
   of the sample galaxies as a function of their morphological type.}
 \label{fig:grad_type_1burst}
\end{figure}

\begin{figure}
 \includegraphics[angle=90,width=0.5\textwidth]{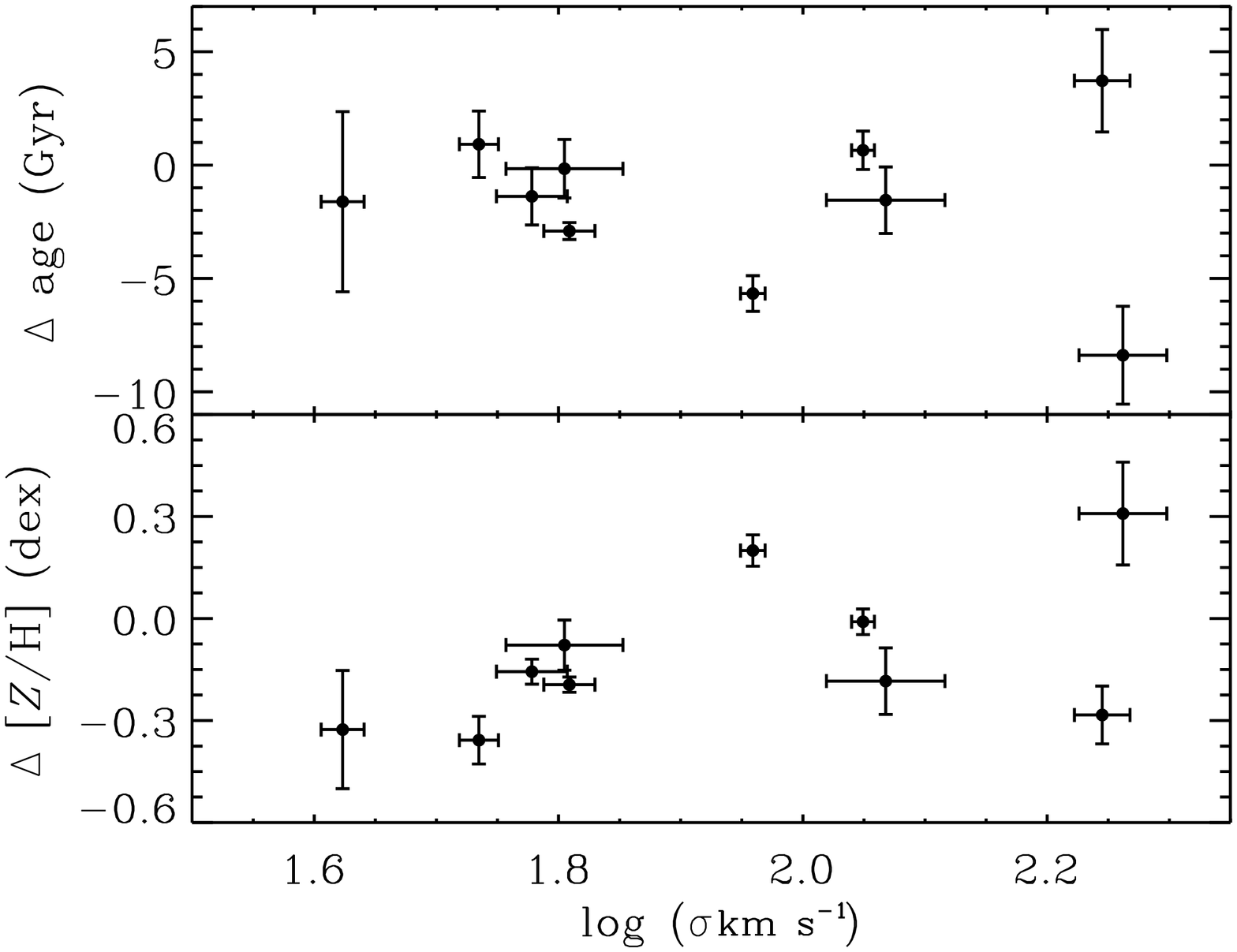}
 \caption{Gradients of the luminosity-weighted age (left-hand panel)
   and metallicity (right-hand panel) of the stellar populations in
   the discs of the sample galaxies as a function of their central
   velocity dispersion. }
 \label{fig:grad_sigma_1burst}
\end{figure}

We note that the age and metallicity gradients do not show any trend
with the galaxy morphological type (Fig.~\ref{fig:grad_type_1burst})
or central velocity dispersion (Fig.~\ref{fig:grad_sigma_1burst})
supporting earlier findings of \citet{sancetal14}. However, the wide
range of masses of the sample galaxies could, in principle, blur the
effect of the morphological type on the correlations with the stellar
population properties. For this reason, these results need to be
tested with a larger sample of galaxies where the effects of the
morphology could be tested for galaxies in a similar mass range.

\subsubsection{Young and old stellar populations}
\label{sec:2burst}

The age distribution obtained by fitting the observed spectra with
synthetic population models is bimodal in most of the sample galaxies.
Some galaxies are also characterised by a bimodal metallicity
distribution (Fig.~\ref{fig:age_metallicity}).

We interpreted such a bimodality being due to the presence of two stellar
populations with a different age. In this paper we assumed the young
stellar population having $\rm Age \leq 4$ Gyr and the old one to
have $\rm Age > 4$ Gyr. We derived their luminosity-weighted age and
metallicity at \Rnc\/ and \Rlast. The errors on age and metallicity
were obtained from photon statistics and CCD readout noise, and they
were calibrated through a series of Monte Carlo simulations. The
values are plotted in Fig.~\ref{fig:age_metallicity} and listed in
Table \ref{tab:results_2burst}.

\begin{figure}
\includegraphics[angle=90.0,width=0.49\textwidth]{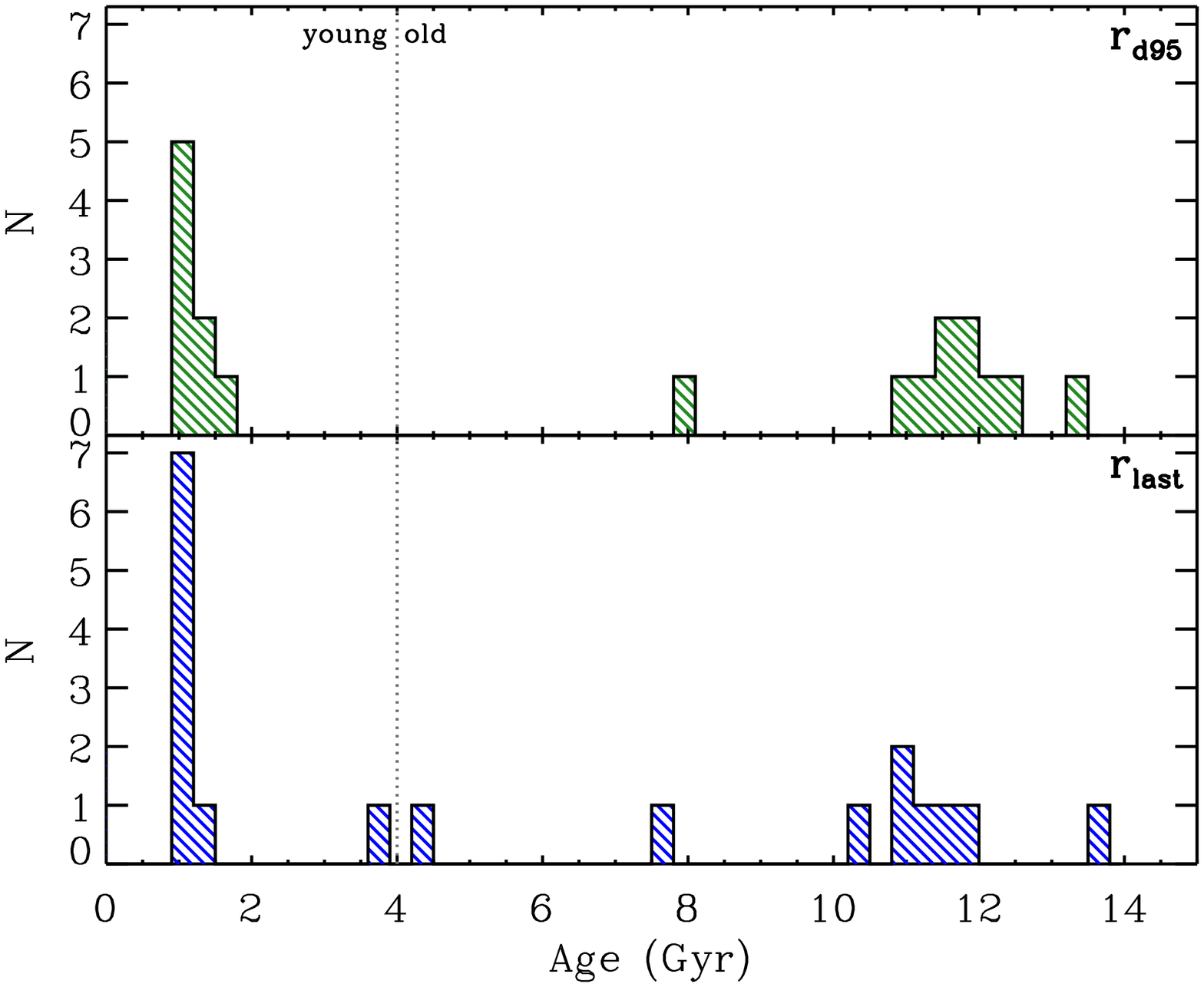}
 \caption{Distribution of the luminosity-weighted age calculated at
   \Rnc\ (green histograms, upper panels) and \Rlast\ (blue
   histograms, lower panels) for the young (left-hand panels) and old
   (right-hand panels) stellar populations in the discs of the sample
   galaxies. The vertical dotted line marks Age$\,=\,4$ Gyr. }
 \label{fig:age_histo_2burst}
\end{figure}

\begin{figure}
\includegraphics[angle=90.0,width=0.49\textwidth]{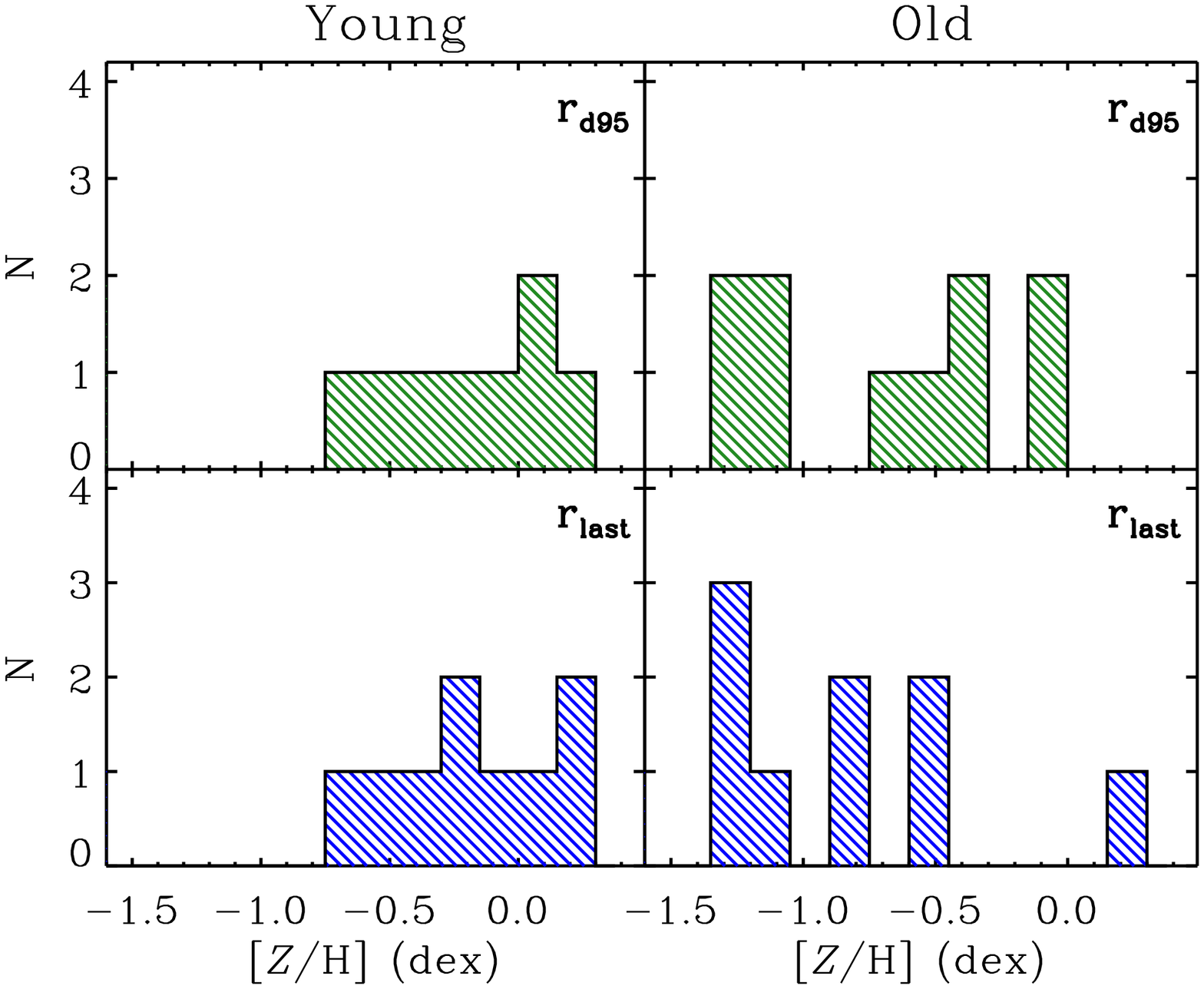}
 \caption{Distribution of the luminosity-weighted metallicity
   calculated at \Rnc\ (green histograms, upper panels) and
   \Rlast\ (blue histograms, lower panels) for the young (left-hand
   panels) and old (right-hand panels) stellar populations in the
   discs of the sample galaxies. }
 \label{fig:met_histo_2burst}
\end{figure}

The number distributions of the luminosity-weighted age and
metallicity of the two stellar populations are shown in
Fig.~\ref{fig:age_histo_2burst} and \ref{fig:met_histo_2burst},
respectively.  The old component covers a large range of metallicities
($-1.3\ltsim\,$\metlumo$\,\ltsim0.1$ dex) at both \Rnc\ and \Rlast,
whereas the metallicity range of the young stellar population is
slightly smaller and shifted towards positive values
($-0.6\ltsim\,$\metlumo$\,\ltsim0.2$ dex).
 
\renewcommand{\tabcolsep}{2pt}
\begin{table*}
\caption{Luminosity-weighted age and metallicity measured for the
  young ($\rm Age \leq 4$ Gyr) and old ($\rm Age > 4$ Gyr) stellar
  populations in the disc-dominated region of the sample
  galaxies. The columns show the following: 1, galaxy name; 2,
  luminosity-weighted age of the young population at \Rnc; 3,
  luminosity-weighted age of the old population at \Rnc; 4,
  luminosity-weighted metallicity of the young population at \Rnc;
  5, luminosity-weighted metallicity of the old population at \Rnc;
  6, luminosity fraction of the old population at \Rnc; 7,
  luminosity-weighted age of the young population at \Rlast; 8,
  luminosity-weighted age of the old population at \Rlast; 9,
  luminosity-weighted metallicity of the young population at \Rlast;
  10, luminosity-weighted metallicity of the old population at
  \Rlast; 11, luminosity fraction of the old population at
  \Rlast. }
\label{tab:results_2burst}
\begin{center}
\begin{tabular}{l rrrr c rrrr c}
\hline
\noalign{\smallskip}
\multicolumn{1}{c}{} & 
\multicolumn{5}{c}{\Rnc} & 
\multicolumn{5}{c}{\Rlast} \\
\multicolumn{1}{c}{Galaxy} & 
\multicolumn{1}{c}{\agelumy} &
\multicolumn{1}{c}{\agelumo} &
\multicolumn{1}{c}{\metlumy} &
\multicolumn{1}{c}{\metlumo} &
\multicolumn{1}{c}{$L_{\rm old}/L_T$} &
\multicolumn{1}{c}{\agelumy} &
\multicolumn{1}{c}{\agelumo} &
\multicolumn{1}{c}{\metlumy} &
\multicolumn{1}{c}{\metlumo} &
\multicolumn{1}{c}{$L_{\rm old}/L_T$} \\
\multicolumn{1}{c}{} & 
\multicolumn{1}{c}{(Gyr)} & 
\multicolumn{1}{c}{(Gyr)} & 
\multicolumn{1}{c}{(dex)} & 
\multicolumn{1}{c}{(dex)} & 
\multicolumn{1}{c}{} &
\multicolumn{1}{c}{(Gyr)} & 
\multicolumn{1}{c}{(Gyr)} & 
\multicolumn{1}{c}{(dex)} & 
\multicolumn{1}{c}{(dex)} & 
\multicolumn{1}{c}{} \\
\multicolumn{1}{c}{(1)} &
\multicolumn{1}{c}{(2)} &
\multicolumn{1}{c}{(3)} &
\multicolumn{1}{c}{(4)} &
\multicolumn{1}{c}{(5)} &
\multicolumn{1}{c}{(6)} &
\multicolumn{1}{c}{(7)} &
\multicolumn{1}{c}{(8)} &
\multicolumn{1}{c}{(9)} &
\multicolumn{1}{c}{(10)} &
\multicolumn{1}{c}{(11)} \\ 
\noalign{\smallskip}
\hline		    
\noalign{\smallskip}        	                		 
ESO-LV~1890070 & $...$       & $11.7\pm0.9$ & $...$          & $-0.45\pm0.03$ & 1.00 & $1.0\pm0.1$ & $13.6\pm1.9$ & $-0.20\pm0.06$ & $ 0.16\pm0.08$ & $0.16$\\ 
ESO-LV~2060140 & $1.1\pm0.4$ & $11.7\pm2.1$ & $-0.01\pm0.08$ & $-1.19\pm0.17$ & 0.29 & $1.1\pm0.2$ & $11.1\pm2.8$ & $-0.64\pm0.16$ & $-1.31\pm0.21$ & $0.42$\\ 
ESO-LV~4000370 & $1.1\pm0.1$ & $12.1\pm2.2$ & $-0.63\pm0.10$ & $-1.30\pm0.05$ & 0.38 & $1.1\pm0.2$ & $ 4.3\pm3.5$ & $ 0.22\pm0.46$ & $-1.28\pm0.07$ & $0.90$\\ 
ESO-LV~4500200 & $1.3\pm0.2$ & $ 8.0\pm3.1$ & $-0.39\pm0.07$ & $-1.30\pm0.18$ & 0.16 & $1.0\pm0.1$ & $10.3\pm2.9$ & $-0.25\pm0.09$ & $-1.31\pm0.08$ & $0.29$\\ 
ESO-LV~5140100 & $1.1\pm0.2$ & $11.0\pm1.5$ & $ 0.14\pm0.12$ & $-0.53\pm0.13$ & 0.70 & $1.2\pm0.1$ & $11.3\pm1.5$ & $-0.42\pm0.11$ & $-0.90\pm0.16$ & $0.42$\\ 
ESO-LV~5480440 & $1.4\pm0.1$ & $11.5\pm1.3$ & $ 0.18\pm0.06$ & $-0.41\pm0.05$ & 0.58 & $3.6\pm0.9$ & $10.9\pm1.5$ & $ 0.22\pm0.24$ & $-1.10\pm0.25$ & $0.46$\\ 
IC~1993        & $...$       & $11.9\pm1.4$ & $...$          & $-0.64\pm0.06$ & 1.00 & $1.3\pm0.2$ & $ 7.7\pm0.5$ & $ 0.08\pm0.09$ & $-0.89\pm0.11$ & $0.46$\\ 
NGC~1366       & $1.1\pm0.3$ & $12.5\pm1.1$ & $-0.57\pm0.10$ & $-0.05\pm0.04$ & 0.56 & $...$       & $11.6\pm1.6$ & $...$          & $-0.56\pm0.06$ & $1.00$\\ 
NGC~7643       & $1.0\pm0.2$ & $13.4\pm1.4$ & $-0.15\pm0.14$ & $-0.10\pm0.07$ & 0.57 & $1.1\pm0.3$ & $11.8\pm1.2$ & $-0.02\pm0.15$ & $-0.59\pm0.11$ & $0.51$\\ 
PGC~37759      & $1.6\pm0.1$ & $11.3\pm1.1$ & $ 0.15\pm0.02$ & $-1.09\pm0.07$ & 0.34 & $1.1\pm0.1$ & $...$        & $-0.54\pm0.02$ & $...$          & $0.00$\\ 
\noalign{\smallskip}
\hline				    	    			 
\end{tabular}
\end{center}
\end{table*}

\begin{figure}
\includegraphics[angle=90.0,width=0.5\textwidth]{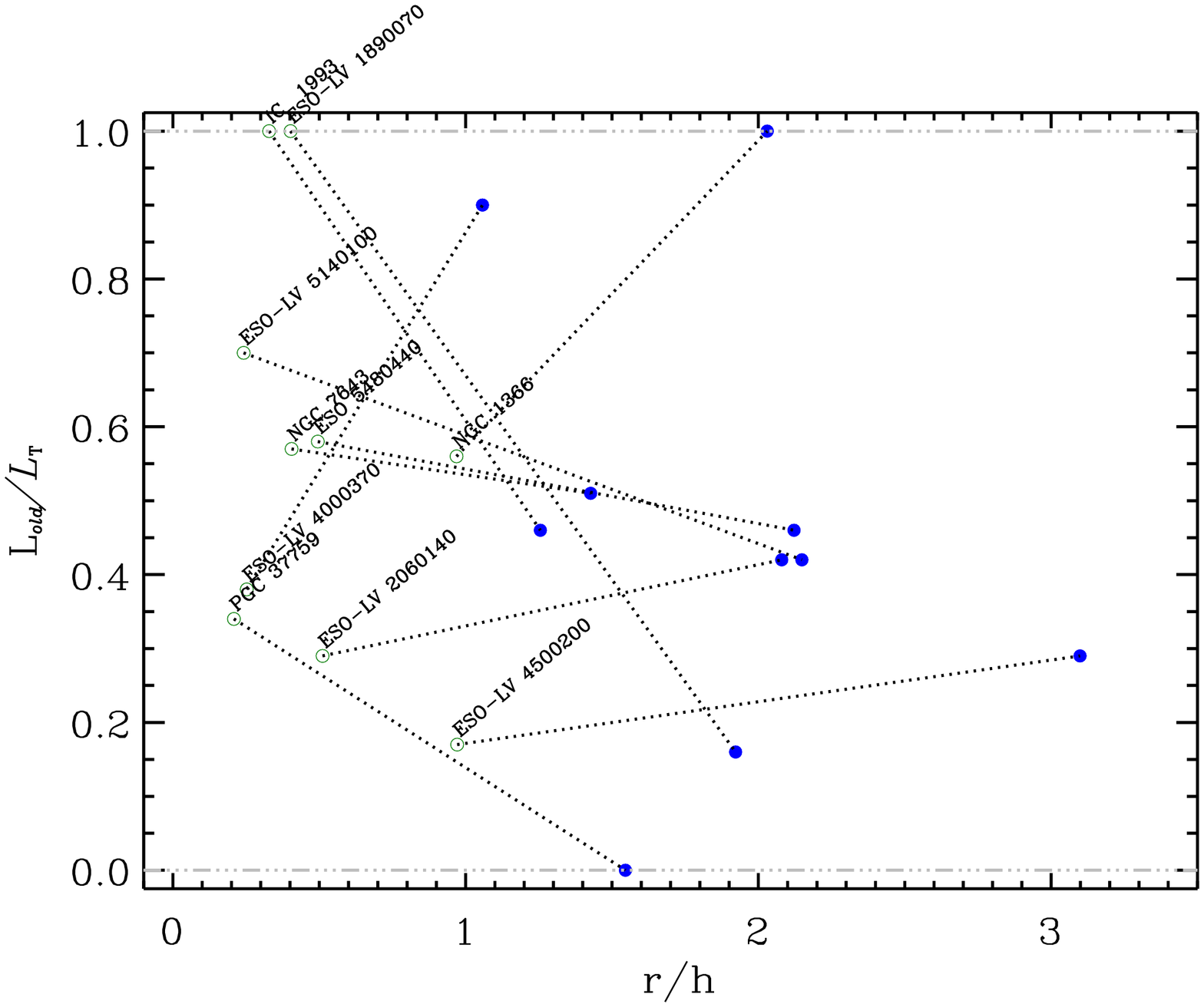}
\caption{Luminosity fraction of the old stellar population at
  \Rnc\ (green open circles) and \Rlast\ (blue circles) in the discs of the
  sample galaxies. The dotted lines connect the values obtained for
  the same galaxy.}
\label{fig:frac_old_2burst}
\end{figure}

The fraction of total luminosity contributed by the old stellar
population at \Rnc\/ and \Rlast\ is given in Table
\ref{tab:results_2burst} and plotted in
Fig.~\ref{fig:frac_old_2burst}. The galaxy luminosity in the
disc-dominated region of about half of the sample galaxies is mostly
contributed by the old stellar population. Its luminosity fraction is
almost constant within the radial range between \Rnc\ and \Rlast\ in
ESO-LV~2060140, ESO-LV~4500200, ESO-LV~5140100, ESO-LV~5480440, and
NGC~7643, whereas it displays a significant change in the remaining
galaxies. The fraction of old stars strongly decreases in the outer
regions of the disc of ESO-LV~1890070, IC~1993, and PGC~37759 and
sharply increases in ESO-LV~400037 and NGC~1366. Therefore no age
gradient is observed in the discs of half of the sample galaxies.

\begin{figure*}
 \includegraphics[angle=90.0,width=0.49\textwidth]{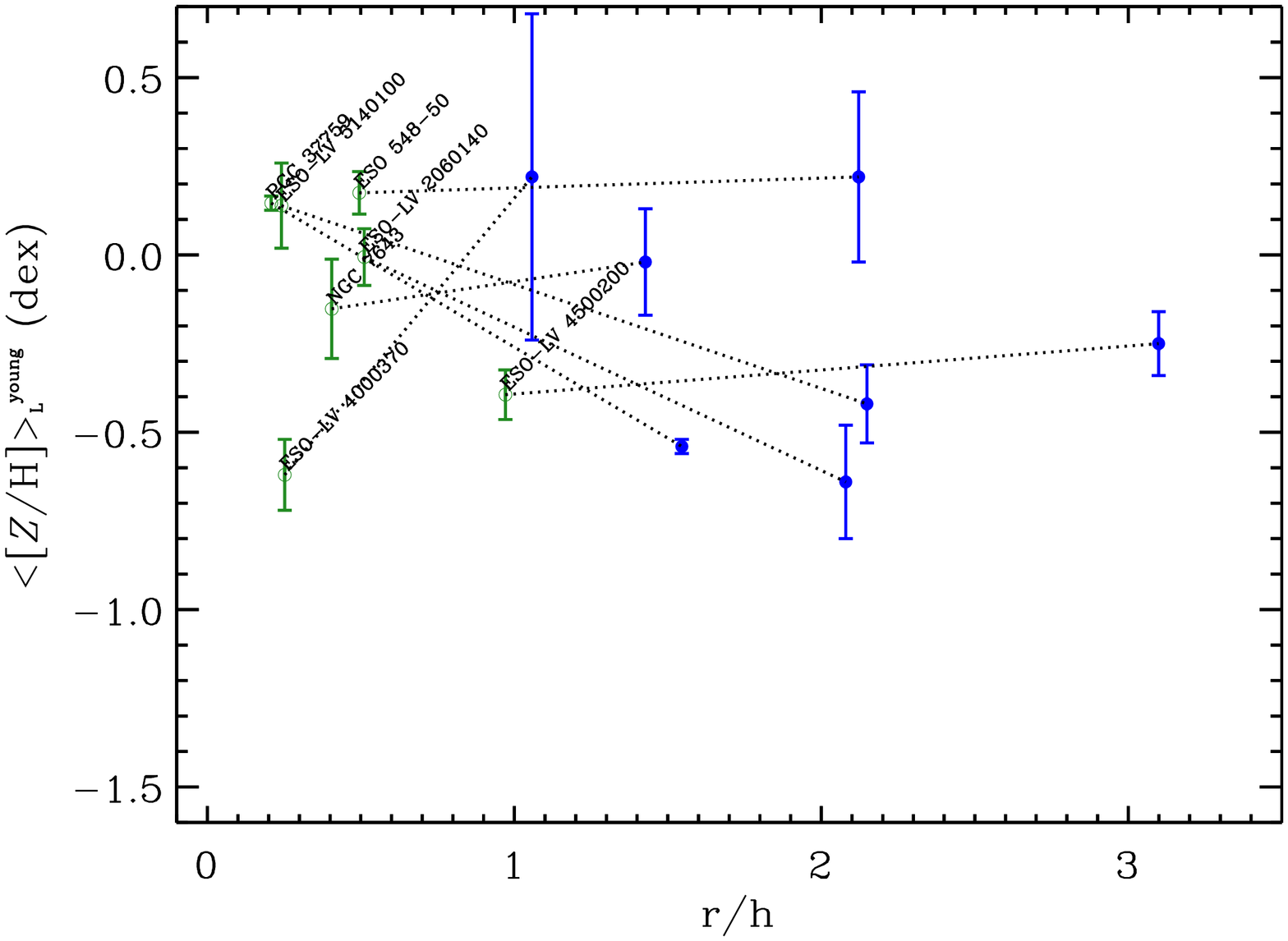}
 \includegraphics[angle=90.0,width=0.49\textwidth]{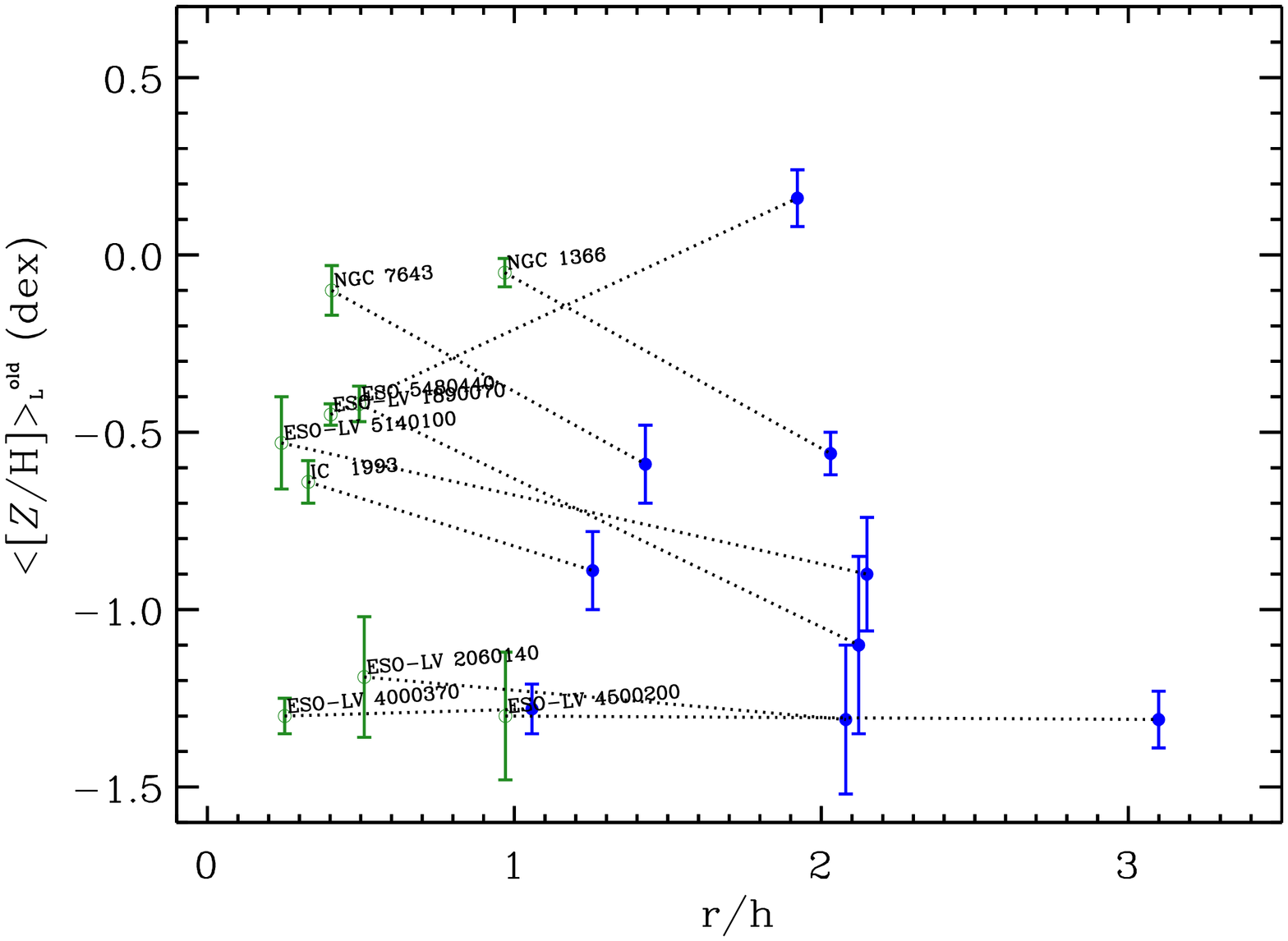}
 \caption{Luminosity-weighted metallicity calculated at \Rnc\ (green
   open circles) and \Rlast\ (blue circles) for the young (left-hand
   panels) and old (right-hand panels) stellar populations in the
   discs of the sample galaxies. The dotted lines connect the values
   obtained for the same galaxy.}
 \label{fig:met_trend_2burst}
\end{figure*}

The luminosity-weighted metallicity measured at \Rnc\/ and \Rlast\ for
the young and old stellar populations are plotted in
Fig.~\ref{fig:met_trend_2burst}. 

\begin{figure}
 \includegraphics[angle=90.0,width=0.49\textwidth]{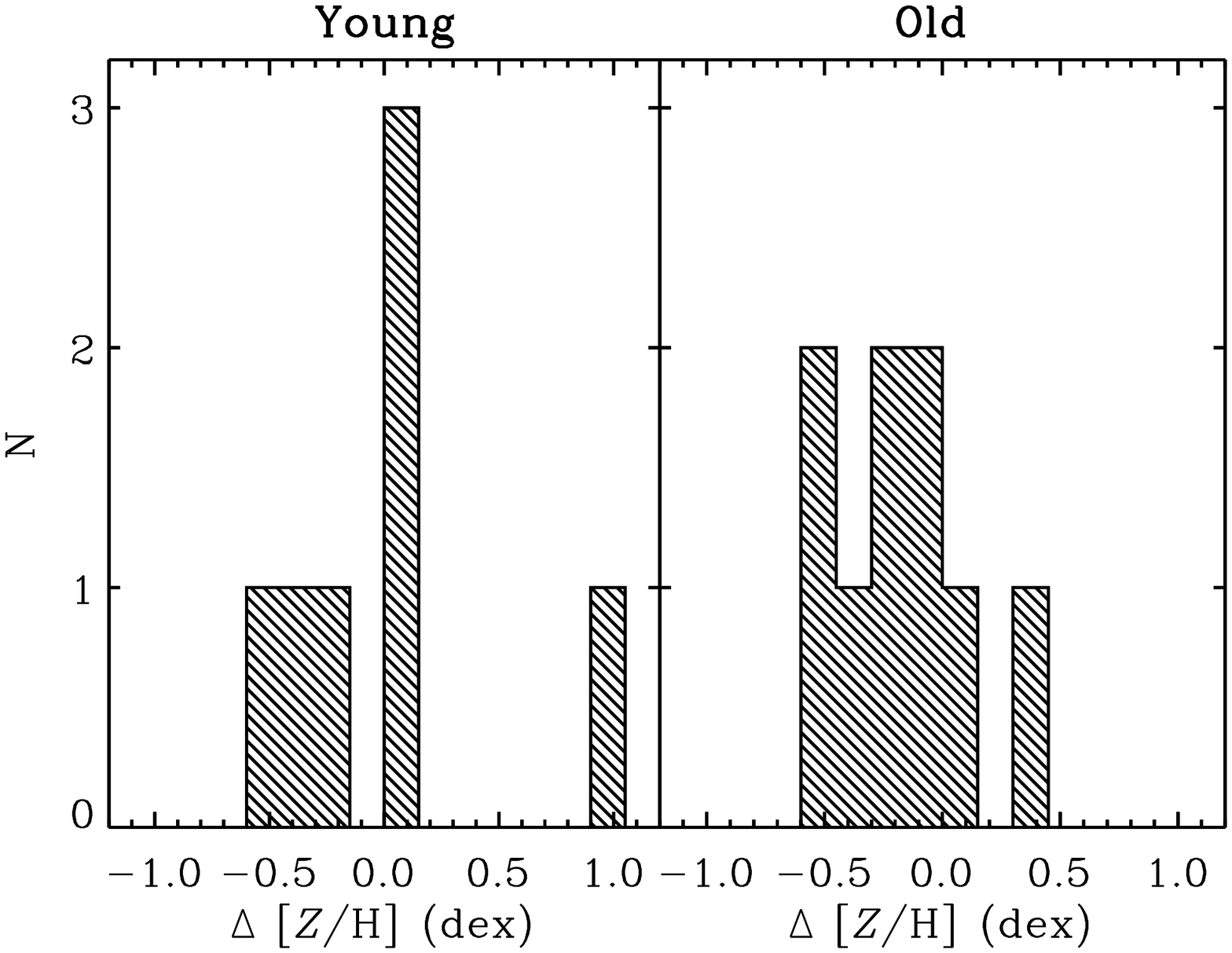}
 \caption{Distribution of the gradients of the luminosity-weighted
   metallicity of the young (left-hand panels) and old (right-hand
   panels) stellar populations in the discs of the sample galaxies.}
 \label{fig:grad_histo_2burst}
\end{figure}

The metallicity gradients over the disc scalelength were derived
separately for the young and old stellar populations. They are
listed in Table~\ref{tab:grad_met_2burst} and their number
distributions are shown in Fig.~\ref{fig:grad_histo_2burst}. 

The distribution of the metallicity gradients of the old stellar
population is similar to that of the mean stellar population with a
prevalence towards negative gradients ($\Delta$\metlumo$\,\simeq-0.22$
dex) indicating that the inner disc regions are more metal rich than
the outer ones. For the young stellar population we found that about
half of the galaxies have slightly positive gradients
($\Delta$\metlumy$\,\simeq0.03$ dex). These results are consistent
with those found by \citet{sancetal14} for the old and young stellar
components of the discs of their galaxies when rescaled to the disc
effective radius.

\begin{table}
\caption{Gradients of metallicity of the young ($\rm Age \leq 4$
  Gyr) and old ($\rm Age > 4$ Gyr) stellar populations measured over
  the disc scalelength in the disc-dominated region of the sample
  galaxies.}
\begin{center}
\begin{small}
\begin{tabular}{lrr}
\hline
\noalign{\smallskip}
\multicolumn{1}{c}{Galaxy} &
\multicolumn{1}{c}{$\Delta$\metlumy} &
\multicolumn{1}{c}{$\Delta$\metlumo} \\
\multicolumn{1}{c}{} &
\multicolumn{1}{c}{(dex)} &
\multicolumn{1}{c}{(dex)} \\
\multicolumn{1}{c}{(1)} &
\multicolumn{1}{c}{(2)} &
\multicolumn{1}{c}{(3)} \\
\noalign{\smallskip}
\hline
\noalign{\smallskip}  
ESO-LV~1890070 & $ ...         $ & $ 0.40\pm0.07$ \\
ESO-LV~2060140 & $-0.40 \pm0.15$ & $-0.08\pm0.24$ \\
ESO-LV~4000370 & $ 1.04 \pm0.69$ & $ 0.02\pm0.14$ \\
ESO-LV~4500200 & $ 0.07 \pm0.07$ & $-0.01\pm0.12$ \\
ESO-LV~5140100 & $-0.29 \pm0.12$ & $-0.19\pm0.15$ \\
ESO-LV~5480440 & $ 0.03 \pm0.18$ & $-0.42\pm0.18$ \\
IC~1993        & $ ...         $ & $-0.27\pm0.18$ \\
NGC~1366       & $ ...         $ & $-0.48\pm0.09$ \\
NGC~7643       & $ 0.13 \pm0.28$ & $-0.48\pm0.17$ \\
PGC~37759      & $-0.51 \pm0.02$ & $...         $ \\
\noalign{\smallskip}
\hline
\end{tabular}
\end{small}
\label{tab:grad_met_2burst}
\end{center}
\end{table}

\subsection{Overabundance of $\alpha$-elements}
\label{sec:ssp}

The overabundance of the $\alpha$-elements over iron is important to
understand the processes driving the formation and evolution of the
galaxies. Indeed it is a proxy of the delay between the supernovae
type II and type I and it gives indication of the timescale of the last
major burst of star formation. We derived the \aFe\ enhancement by
comparing the measurements of line-strength indices with the
predictions of SSP models.

\subsubsection{Measuring the line-strength indices}
\label{sec:indices}

\citet{pizzetal08} and \citet{moreetal08, morelli2012, morelli15}
measured the radial profiles of the Lick \Hb , Mg, and Fe
line-strength indices out to $2-3\,h$ from the centre for all the
sample galaxies. The values of the \Hb\ line-strength index were
corrected for the possible contamination of the \Hb\ emission line due
to the ionized-gas component \citep[see][for details]{morelli2012}.

We derived the line-strength indices at \Rnc\/ and \Rlast\ by linearly
interpolating the measured line-strength indices along the radius. The
resulting values and their corresponding errors are reported in
Table~\ref{tab:indices}.

\renewcommand{\tabcolsep}{5pt}
\begin{table*}
\caption{Values of the line-strength indices of the sample galaxies
  measured at \Rnc\/ and \Rlast. }
\begin{tiny}
\begin{tabular}{l ccccc ccrcc}
\hline
\noalign{\smallskip}
\multicolumn{1}{c}{} &
\multicolumn{5}{c}{\Rnc} &
\multicolumn{5}{c}{\Rlast} \\
\multicolumn{1}{c}{Galaxy} &
\multicolumn{1}{c}{\Fe} &
\multicolumn{1}{c}{\MgFe} &
\multicolumn{1}{c}{\Mgd} &
\multicolumn{1}{c}{\Mgb} &
\multicolumn{1}{c}{\Hb} &
\multicolumn{1}{c}{\Fe} &
\multicolumn{1}{c}{\MgFe} &
\multicolumn{1}{c}{\Mgd} &
\multicolumn{1}{c}{\Mgb} &
\multicolumn{1}{c}{\Hb} \\
\multicolumn{1}{c}{} &
\multicolumn{1}{c}{(\AA)} &
\multicolumn{1}{c}{(\AA)} &
\multicolumn{1}{c}{(mag)} &
\multicolumn{1}{c}{(\AA)} &
\multicolumn{1}{c}{(\AA)} & 
\multicolumn{1}{c}{(\AA)} &
\multicolumn{1}{c}{(\AA)} &
\multicolumn{1}{c}{(mag)} &
\multicolumn{1}{c}{(\AA)} &
\multicolumn{1}{c}{(\AA)} \\
\multicolumn{1}{c}{(1)} &
\multicolumn{1}{c}{(2)} &
\multicolumn{1}{c}{(3)} &
\multicolumn{1}{c}{(4)} &
\multicolumn{1}{c}{(5)} &
\multicolumn{1}{c}{(6)} &
\multicolumn{1}{c}{(7)} &
\multicolumn{1}{c}{(8)} &
\multicolumn{1}{c}{(9)} &
\multicolumn{1}{c}{(10)} &
\multicolumn{1}{c}{(11)} \\ 
\noalign{\smallskip}
\hline
\noalign{\smallskip}
ESO-LV~1890070 & $2.37\pm0.13$ & $2.81\pm0.09$ & $0.191\pm0.004$ & $3.17\pm0.09$ & $ 1.53\pm0.09$ & $2.05\pm0.09$ & $2.30\pm0.10$ & $ 0.124\pm0.004$ & $2.47\pm0.04$ & $3.61\pm0.05$ \\
ESO-LV~2060140 & $1.68\pm0.19$ & $1.82\pm0.13$ & $0.088\pm0.006$ & $1.91\pm0.11$ & $ 3.51\pm0.11$ & $0.92\pm0.28$ & $1.27\pm0.17$ & $ 0.070\pm0.006$ & $1.47\pm0.13$ & $3.55\pm0.16$ \\
ESO-LV~4000370 & $1.26\pm0.19$ & $1.27\pm0.13$ & $0.061\pm0.006$ & $1.30\pm0.12$ & $ 3.45\pm0.11$ & $1.26\pm0.28$ & $1.17\pm0.18$ & $ 0.031\pm0.006$ & $1.14\pm0.17$ & $3.32\pm0.17$ \\
ESO-LV~4500200 & $1.63\pm0.10$ & $1.65\pm0.06$ & $0.097\pm0.005$ & $1.66\pm0.06$ & $ 3.30\pm0.05$ & $1.59\pm0.15$ & $1.61\pm0.12$ & $ 0.103\pm0.005$ & $1.60\pm0.08$ & $3.50\pm0.09$ \\
ESO-LV~5140100 & $2.10\pm0.10$ & $2.22\pm0.07$ & $0.141\pm0.003$ & $2.25\pm0.06$ & $ 1.94\pm0.06$ & $1.29\pm0.15$ & $1.57\pm0.08$ & $ 0.102\pm0.003$ & $1.89\pm0.09$ & $3.09\pm0.08$ \\
ESO-LV~5480440 & $2.26\pm0.12$ & $2.38\pm0.09$ & $0.147\pm0.010$ & $2.55\pm0.08$ & $ 1.88\pm0.07$ & $2.16\pm0.11$ & $2.27\pm0.07$ & $ 0.139\pm0.006$ & $2.30\pm0.07$ & $2.48\pm0.07$ \\
IC  1993       & $2.09\pm0.13$ & $2.34\pm0.08$ & $0.148\pm0.008$ & $2.56\pm0.07$ & $ 1.85\pm0.06$ & $2.19\pm0.13$ & $2.48\pm0.09$ & $ 0.148\pm0.012$ & $2.80\pm0.07$ & $2.74\pm0.06$ \\
NGC 1366       & $2.35\pm0.22$ & $2.66\pm0.14$ & $0.187\pm0.015$ & $3.05\pm0.13$ & $ 1.73\pm0.10$ & $1.99\pm0.14$ & $2.44\pm0.08$ & $ 0.171\pm0.012$ & $2.96\pm0.08$ & $1.50\pm0.06$ \\
NGC 7643       & $2.51\pm0.20$ & $2.80\pm0.13$ & $0.179\pm0.017$ & $2.98\pm0.12$ & $ 2.24\pm0.12$ & $2.12\pm0.16$ & $2.35\pm0.11$ & $ 0.143\pm0.006$ & $2.58\pm0.11$ & $3.08\pm0.09$ \\
PGC~37759      & $2.18\pm0.23$ & $2.56\pm0.14$ & $0.057\pm0.002$ & $2.93\pm0.11$ & $ 2.97\pm0.11$ & $1.51\pm0.22$ & $1.80\pm0.14$ & $-0.013\pm0.007$ & $1.99\pm0.11$ & $4.53\pm0.10$ \\
\noalign{\smallskip}
\hline
\label{tab:indices}
\end{tabular}
\end{tiny}
\end{table*}

\subsubsection{Total [$\alpha/$Fe] enhancement}
\label{sec:alfe}

\begin{figure}
\centering
\includegraphics[angle=90.0,width=0.495\textwidth]{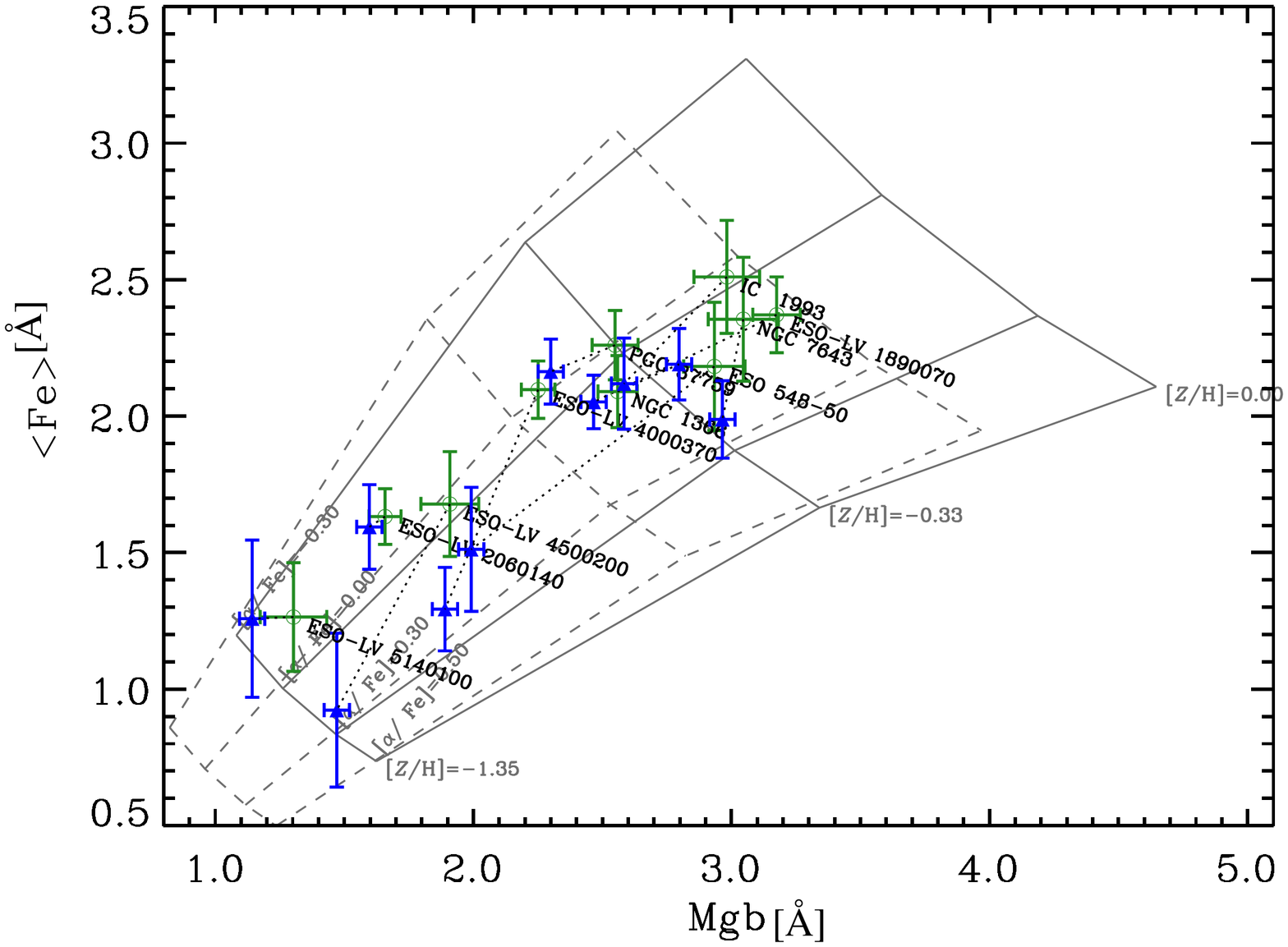}
\caption{Distribution of the \Fe\ and \Mgb\/ line-strength indices
  measured at \Rnc\/ (green open circles) and \Rlast\/ (blue triangles) for
  the sample galaxies. The dotted lines connect the values referred to
  the same galaxy. The light grey lines indicate the models by
  \citet{thmabe03} for a young (4 Gyr, continuous lines) and an old
  (10 Gyr, dashed lines) stellar population. }
\label{fig:models_ssp}
\end{figure}

In Fig.~\ref{fig:models_ssp} the values of \Mgb\ and \Fe\ derived at
\Rnc\/ and \Rlast\/ for each sample galaxy are compared with the model
predictions by \citet{thmabe03} for two stellar populations with an
intermediate (2 Gyr) and old age (12 Gyr), respectively.

\begin{figure}
\centering
\includegraphics[angle=90,width=0.48\textwidth]{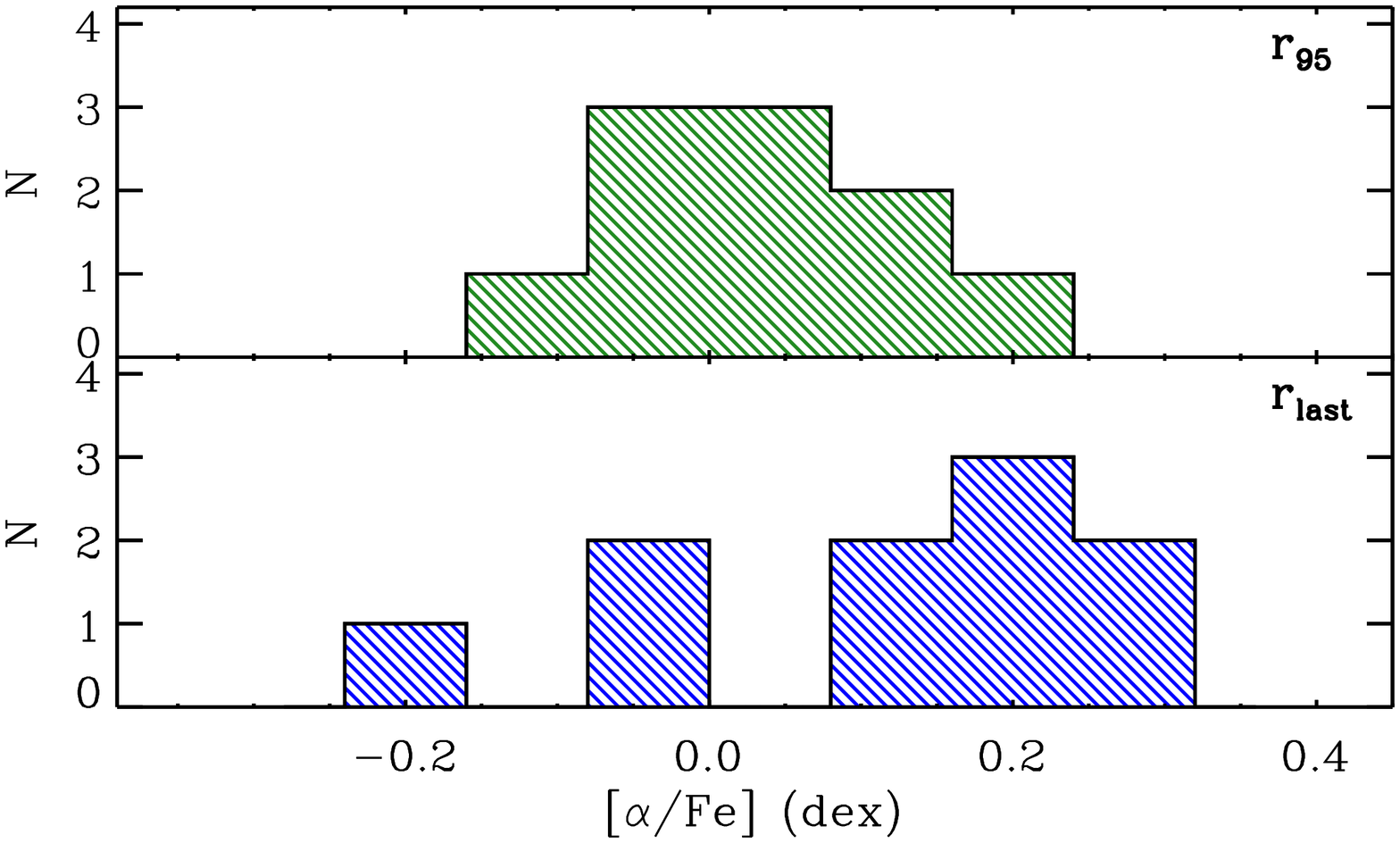}\\
\caption[]{Distribution of the total $\alpha$/Fe enhancement
  calculated at \Rnc\ (green histogram, upper panel) and \Rlast\ (blue
  histogram, lower panel) for the stellar populations in the discs of
  the sample galaxies.}
\label{fig:alfe_histo}
\end{figure}

The total $\alpha/$Fe enhancement of the disc stellar population at
\Rnc\/ and \Rlast\/ was derived from the values of line-strength
indices of Table~\ref{tab:indices}, using a linear interpolation
between the model points with the iterative procedure described in
\citet{moreetal08} and adopting the age given in Table
\ref{tab:results_1burst}.  The uncertainties on the $\alpha/$Fe
enhancements were estimated by Monte Carlo simulations as done in
\cite{moreetal08, morelli15}. We randomly generated 100 simulated sets
of line-strength indices from the measured indices and their errors
assuming Gaussian distributions. The standard deviations of the
distributions of simulated $\alpha/$Fe enhancements were adopted as
the errors on their measured values, which are reported in Table
\ref{tab:results_1burst}.  The histograms of the number distribution
of the \aFe\ enhancement at \Rnc\ and \Rlast\ are plotted in
Fig.~\ref{fig:alfe_histo}. Most of the galaxies display a solar and
super-solar \aFe\ enhancement in the inner and outer regions of the
disc, respectively.

\begin{figure}
\centering
\includegraphics[angle=90,width=0.48\textwidth]{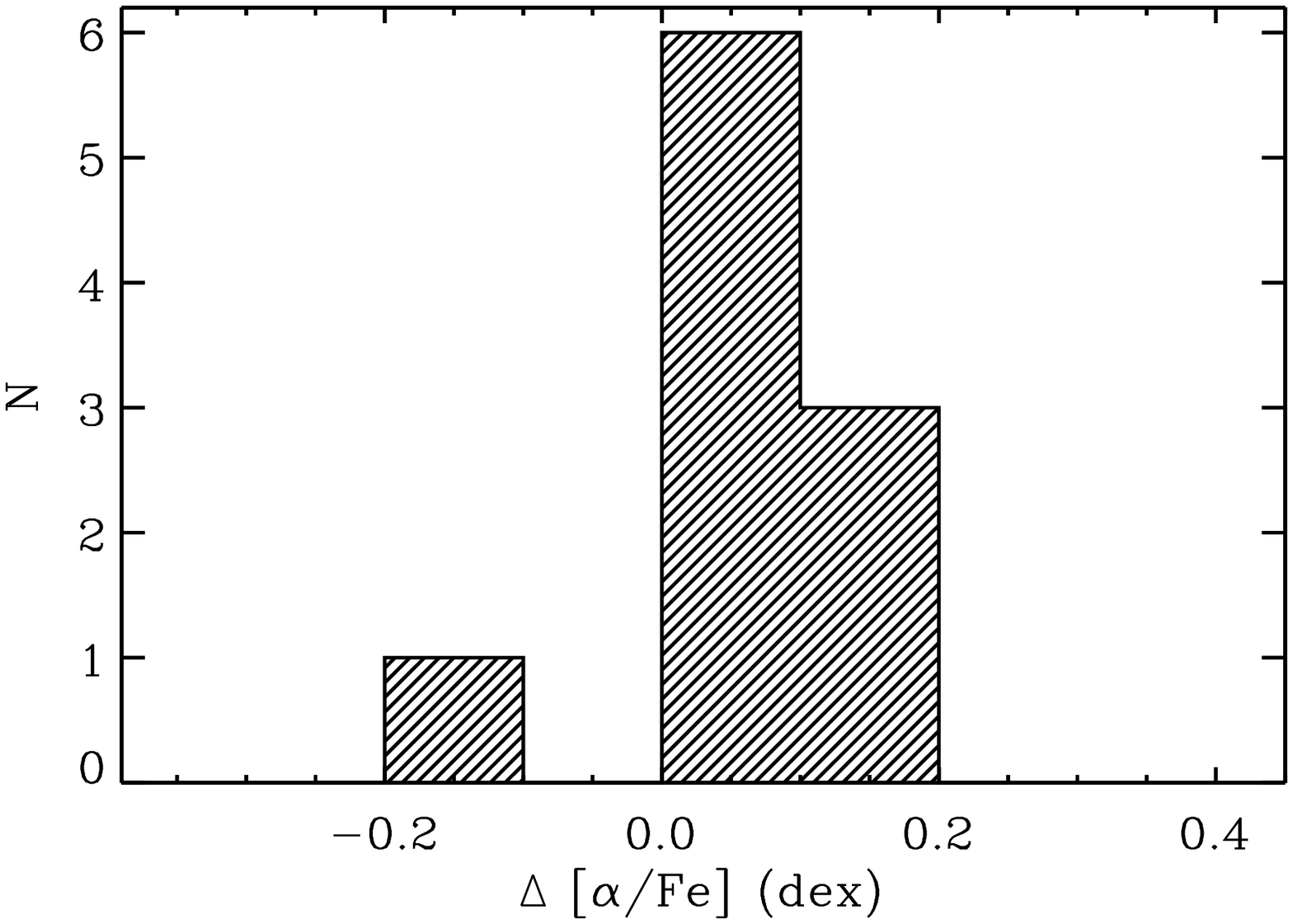}\\
 \caption{Distribution of the gradients of the total \aFe\ enhancement
   of the stellar populations in the discs of the sample galaxies.}
\label{fig:grad_alfe_histo}
\end{figure}

The difference in the distributions of total \aFe\ enhancement
measured at \Rnc\/ and \Rlast\/ suggests the presence of a radial
gradient in the stellar populations properties of some of the
galaxies, as also shown by the trends of Fig.~\ref{fig:models_ssp}.
The gradients of total \aFe\ enhancement over the disc scalelength
were derived for all the sample galaxies from the values measured at
\Rnc\ and \Rlast. The errors on the gradients were calculated through
Monte Carlo simulations as done in \citet{morelli2012}. The gradients
and corresponding errors are listed in Table \ref{tab:alfe} and their
number distribution is shown in Fig.~\ref{fig:grad_alfe_histo}.  Only
the discs of ESO-LV~5140100 and IC~1993 are characterised by a shallow
positive gradient of \aFe\ enhancement, whereas the discs of all the
other galaxies show a \aFe\ gradient consistent with zero within the
errors.  It is worth noticing that all the \aFe\ gradients are
systematically positive with a weighted mean value of
$\overline{[\alpha/{\rm Fe}]}=0.10\pm0.09$ dex.
 
\begin{table}
\caption{Overabundance of the $\alpha$-elements over iron of the
  stellar populations in the disc-dominated region of the sample
  galaxies. The columns show the following: 1, galaxy name; 2, total
  \aFe\ enhancement at \Rnc ; 3, total \aFe\ enhancement at \Rlast ;
  4, gradient of the total \aFe\ enhancement in the disc.}
\begin{center}
\begin{small}
\begin{tabular}{lrrr}
\hline
\noalign{\smallskip}
\multicolumn{1}{c}{} &
\multicolumn{1}{c}{\Rnc} &
\multicolumn{1}{c}{\Rlast}&
\multicolumn{1}{c}{} \\
\multicolumn{1}{c}{Galaxy} &
\multicolumn{1}{c}{\aFe} &
\multicolumn{1}{c}{\aFe} &
\multicolumn{1}{c}{ $\Delta$\aFe}\\
\multicolumn{1}{c}{} &
\multicolumn{1}{c}{(dex)} &
\multicolumn{1}{c}{(dex)} &
\multicolumn{1}{c}{(dex)} \\
\multicolumn{1}{c}{(1)} &
\multicolumn{1}{c}{(2)} &
\multicolumn{1}{c}{(3)} &
\multicolumn{1}{c}{(4)} \\
\noalign{\smallskip}
\hline
\noalign{\smallskip}  
ESO-LV~1890070 & $ 0.11\pm0.07$  & $ 0.17\pm0.04$& $ 0.06\pm0.09$\\
ESO-LV~2060140 & $ 0.09\pm0.05$  & $ 0.00\pm0.12$& $ 0.16\pm0.19$\\
ESO-LV~4000370 & $ 0.09\pm0.09$  & $ 0.25\pm0.16$& $-0.14\pm0.20$\\
ESO-LV~4500200 & $-0.02\pm0.06$  & $-0.02\pm0.06$& $ 0.00\pm0.09$\\
ESO-LV~5140100 & $-0.08\pm0.05$  & $ 0.23\pm0.10$& $ 0.31\pm0.12$\\
ESO-LV~5480440 & $-0.03\pm0.05$  & $-0.02\pm0.06$& $ 0.01\pm0.08$\\
IC~1993        & $ 0.05\pm0.06$  & $ 0.16\pm0.06$& $ 0.11\pm0.09$\\
NGC~1366       & $ 0.07\pm0.10$  & $ 0.08\pm0.07$& $ 0.01\pm0.12$\\
NGC~7643       & $ 0.04\pm0.07$  & $ 0.13\pm0.08$& $ 0.09\pm0.11$\\
PGC~37759      & $ 0.21\pm0.10$  & $ 0.25\pm0.11$& $ 0.04\pm0.15$\\
\noalign{\smallskip}
\hline
\end{tabular}
\end{small}
\label{tab:alfe}
\end{center}
\end{table}

\section{Conclusions}
\label{sec:conclusions}

We derived the stellar population properties in the discs of 10 spiral
galaxies to investigate their assembly history by testing the
predictions of theoretical models and numerical simulations.
To this aim we analysed the galaxy spectra obtained in the radial
range between \Rnc\/ and \Rlast, which are the radii where the disc
contributes more than $95\%$ of the galaxy surface brightness and the
farthest measured radius, respectively. On average such a radial range
extends out to $\sim2$ times the disc scalelength $h$ and it is
$\sim1.5h$ wide.

The luminosity-weighted age \agelum\ and luminosity-weighted
metallicity \metlum\ of the stellar populations were measured at
\Rnc\/ and \Rlast\/ by fitting the galaxy spectra with a linear
combination of stellar population synthesis models. 

The disc stellar population of the sample galaxies has a flat
distribution of \agelum\ ranging from $\sim1$ Gyr to $\sim12$ Gyr at
both \Rnc\/ and \Rlast, however we note that at \Rlast only one galaxy
has an age greater than $\sim8$ Gyr while at \Rnc\/ they are four.
The luminosity-weighted metallicities span a wide range of values from
solar to sub-solar ($-1.2\,\leq\,$\metlum$\,\leq0\,$ dex) but the
number distribution at \Rlast\ is slightly shifted towards lower
metallicities and the peak moves from \metlum$\,\simeq-0.4$ dex at
\Rnc\/ at about \metlum$\,\simeq-0.5$ dex at \Rlast.
The disc stellar populations of the majority of the sample galaxies
are characterised by a negligible \agelum\ gradient over the disc
scalelength and negative \metlum\ gradient, giving observational
support to the inside-out formation scenario \citep{matteucci89,
  roskar08}.
No correlation was found between the galaxy morphological type and
either \agelum\ or \metlum\ and between the morphological type and gradients
of \agelum\ and \metlum.  Even though the small number statistics does not
allow us to trace a firm conclusion, this might suggest that the star
formation in discs is not strongly connected with the galaxy type, as
already pointed out for bulges. Indeed \citet{thda06} and
\citet{morelli2012} found that the evolution of bulges and discs do
not have a strong interplay and they follow independent paths of star
formation.

Most of the discs display a bimodal age distribution hosting a young
($\rm Age \leq 4$ Gyr) and an old ($\rm Age > 4$ Gyr) stellar
population, for which we derived the value and gradient of both the
luminosity-weighted age and metallicity. The old stellar component
usually dominates the disc luminosity and it is slightly more metal
poor than the young stellar component. The luminosity fraction of the
old stellar component is almost constant within the observed radial
range in half of the sample galaxies and therefore no age gradient is
observed in their discs.
The old and young stellar populations are characterised by negative
($\Delta$\metlumo$\,\simeq-0.22$ dex) and slightly positive
($\Delta$\metlumy$\,\simeq0.03$ dex) gradients of metallicity,
respectively. This is in agreement with the findings by
\citet{sancetal14} when rescaled to the disc effective radius.
These results suggest that the discs formed out with a shallow
gradient of age and metallicity, and this is consistent with the
predictions of the inside-out assembly scenario
\citep{Pilkington2012}. The young stellar population could be the
result of a second burst of star formation due to the acquisition of
gas from the environment. This give rise to the homogeneously-mixed
stellar population we observe all over the disc. On the other hand, it
is hard to explain the metallicity gradients of the old stellar
populations in the framework of radial migration, which is expected to
erase the gradients of the stellar population properties by moving
stars from the inner to the outer regions of the disc
\citep{roskar08}. These results suggest a reduced impact of radial
migration on the stellar populations properties.

The overabundance of the $\alpha$-elements over iron were derived at
\Rnc\/ and \Rlast\ by comparing the measurements of line-strength
indices with the predictions of SSP models. 

The gradients of total \aFe\ enhancement calculated over the disc
scalelength are systematically positive ($\overline{[\alpha/{\rm
      Fe}]}=0.10\pm0.09$ dex). This is a hint that the star-formation
timescale is shorter in the outer regions of discs. This result is
promising but it should be tested against a larger sample of galaxy
discs and when the new $\alpha$-enhanced synthetic population models
\citep{vazdekis15} will be released.

\section*{Acknowledgments}
This investigation was based on observations made with ESO Telescopes
at the La Silla Paranal Observatory under programmes 76.B-0375, and
80.B-00754.  This work was supported by Padua University through
grants 60A02-4807/12, 60A02-5857/13, 60A02-5833/14 and CPDA133894. LM
acknowledges financial support from Padua University grant
CPS0204. JMA acknowledges support from the European Research Council
Starting Grant (SEDmorph, P.I. V. Wild)


\bsp

\label{lastpage}

\end{document}